\documentclass[twocolumn,trackchanges]{aastex63}

\newcounter{magicrownumbersCP}
\newcounter{magicrownumbersvetoed}
\newcounter{magicrownumbersvetoedfull}
\newcounter{magicrownumbersCFP}
\newcommand\rownumber{\stepcounter{magicrownumbersCP}\arabic{magicrownumbersCP}}
\newcommand\rownumbervetoed{\stepcounter{magicrownumbersvetoed}\arabic{magicrownumbersvetoed}}
\newcommand\rownumbervetoedfull{\stepcounter{magicrownumbersvetoedfull}\arabic{magicrownumbersvetoedfull}}

\usepackage{tabularx}

\usepackage{bbding}
\usepackage{longtable}
\usepackage{microtype}
\usepackage{graphicx,subfigure}
\usepackage{booktabs} 
\usepackage{enumitem}
\usepackage{amsmath}
\usepackage{multirow}
\usepackage{makecell}
\usepackage{slashbox}
\usepackage[linesnumbered, ruled, vlined]{algorithm2e}
\usepackage{algorithmic}
\newcolumntype{Y}{>{\centering\arraybackslash}X}

\usepackage{array,etoolbox}
\setitemize{leftmargin=10pt,noitemsep,topsep=0pt,parsep=0pt,partopsep=0pt}
\usepackage{hyperref}

\usepackage[flushleft, para]{threeparttable}
\usepackage{booktabs}
\usepackage{array}

\usepackage{scrextend}

\usepackage{comment}

\SetKwInput{KwInput}{Input}
\SetKwInput{KwOutput}{Output}

\newcommand{\kepler}{Kepler}

\newcommand{\vespa}{\texttt{vespa}} 
\newcommand{\ExoMiner}{\texttt{ExoMiner}}
\newcommand{\AstroNet}{\texttt{AstroNet}}  
\newcommand{\ExoNet}{\texttt{ExoNet}}
\newcommand{\GPC}{\texttt{GPC}}
\newcommand{\RFC}{\texttt{RFC}}

\newcommand{\Robovetter}{\texttt{Robovetter}}
\newcommand{\Autovetter}{\texttt{Autovetter}}

\shorttitle{ExoMiner: A Highly Accurate and Explainable Deep Learning Classifier to Mine Exoplanets}
\shortauthors{Valizadegan et. al.}

\begin{document}

\title{Multiplicity Boost Of Transit Signal Classifiers: \\Validation of 69 New Exoplanets Using The Multiplicity Boost of ExoMiner\footnote{This work is dedicated to Women, Life, Freedom movement in Iran.}}

\correspondingauthor{Hamed Valizadegan}
\email{hamed.valizadegan@nasa.gov}

\author[0000-0001-6732-0840]{Hamed Valizadegan}
\affiliation{Universities Space Research Association (USRA), Mountain View, CA 94043, USA}
\affiliation{NASA Ames Research Center (NASA ARC), Moffett Field, CA 94035, USA}

\author[0000-0002-2188-0807]{Miguel J. S. Martinho}
\affiliation{Universities Space Research Association (USRA), Mountain View, CA 94043, USA}
\affiliation{NASA Ames Research Center (NASA ARC), Moffett Field, CA 94035, USA}

\author[0000-0002-4715-9460]{Jon M. Jenkins}
\affiliation{NASA Ames Research Center (NASA ARC), Moffett Field, CA 94035, USA}

\author[0000-0003-1963-9616]{Douglas A. Caldwell}
\affiliation{The SETI Institute, Mountain View, CA  94043, USA}
\affiliation{NASA Ames Research Center (NASA ARC), Moffett Field, CA 94035, USA}

\author[0000-0002-6778-7552]{Joseph D. Twicken}
\affiliation{The SETI Institute, Mountain View, CA  94043, USA}
\affiliation{NASA Ames Research Center (NASA ARC), Moffett Field, CA 94035, USA}

\author[0000-0003-0081-1797]{Stephen T. Bryson}
\affiliation{NASA Ames Research Center (NASA ARC), Moffett Field, CA 94035, USA}



\begin{abstract}
Most existing exoplanets are discovered using validation techniques rather than being confirmed by complementary observations. These techniques generate a score that is typically the probability of the transit signal being an exoplanet (y(x)=exoplanet) given some information related to that signal (represented by x). Except for the validation technique in~\cite{Rowe_2014_multiplanet} that uses multiplicity information to generate these probability scores, the existing validation techniques ignore the multiplicity boost information. In this work, we introduce a framework with the following premise: given an existing transit signal vetter (classifier), improve its performance using multiplicity information. We apply this framework to several existing classifiers, which include \vespa~\citep{Morton-2016-vespa},  \Robovetter~\citep{Coughlin2017robovetter}, \AstroNet~\citep{shallue_2018}, \ExoNet~\citep{Ansdell_2018}, \GPC\ and \RFC~\citep{armstrong-2020-exoplanet}, and \ExoMiner~\citep{Valizadegan_2022_ExoMiner}, to support our claim that this framework is able to improve the performance of a given classifier. We then use the proposed multiplicity boost framework for \ExoMiner\ V1.2, which addresses some of the shortcomings of the original \ExoMiner\ classifier~\citep{Valizadegan_2022_ExoMiner}, and validate 69 new exoplanets for systems with multiple KOIs from the Kepler catalog.
\end{abstract}

\section{Introduction}
\label{introduction}

As the traditional approach for the confirmation of new exoplanets, which requires complementary observations, is not possible or practical for all candidates due to the increase in the number of candidates and their specifics (e.g., small planets around faint stars), the focus of the discovery of new exoplanets has been shifting from manual follow-up studies to mass validation using automated processes. Such automated processes include statistical techniques~\citep{Rowe_2014_multiplanet, Morton-2016-vespa} or machine learning based approaches~\citep{shallue_2018, armstrong-2020-exoplanet, Valizadegan_2022_ExoMiner}. 

The statistical approaches include \vespa~\citep{Morton-2016-vespa}, a generative (Bayesian) approach~\citep{Vapnik-statisticallearningtheory-1998} to calculate the posterior probability of a transit signal for classes of interest (planet candidate or different false positive scenarios), and multiplicity boost~\citep{Rowe_2014_multiplanet}, a frequentist approach that calculates the probability of a specific planet and False Positive (FP) configuration (i.e., a specific number of planets and FPs) in a multi-planet system. The generative approach to classification models the prior and likelihood in order to obtain the posterior probability. This is in contrast to the discriminative approach, introduced by \citet{Vapnik-statisticallearningtheory-1998}, that models the posterior probability directly without the need to model priors and likelihood. The compelling justification of \citet{Vapnik-statisticallearningtheory-1998} to propose the discriminative approach was that, ``one should solve the [classification] problem directly and never solve a more general problem as an intermediate step [such as modeling $p(X|y)$].'' 

The machine learning approaches applied to transiting planet data mainly consist of discriminative approaches~\citep{Vapnik-statisticallearningtheory-1998} to calculate the posterior probability and include the following methods: 1) Autovetter~\citep{Jenkins-Autovetter-2014IAUS,McCauliff_2015}, a random forest to classify transit signals summarized in the form of diagnostic metrics represented by scalar values, 2) the first Deep Neural Network (DNN) model to classify transit signals, \AstroNet~\citep{shallue_2018}, that was used to validate two new exoplanets, 3) the machine classifiers developed in~\citet{armstrong-2020-exoplanet} that were used to validate 50 new exoplanets, and 4) a most recently developed DNN model called \ExoMiner~\citep{Valizadegan_2022_ExoMiner} that validated 301 new exoplanets. These statistical and machine learning models validate new exoplanets when the model's confidence about the Planet Candidate (PC) disposition of a transit signal is sufficiently high (typically $>0.99$). 

Most existing validation models rely on the posterior probability of a transit signal alone for validation; i.e., they do not take into account the configuration of a target star and whether there are existing known planets or FPs in that system. However, knowledge of existing planets and FPs for a given system should affect our confidence about the validation of a new unknown signal. More importantly, it is strongly believed that the candidates in multi-planet systems are highly likely planets~\citep{Latham_2011_multiplicity, Lissauer_2012_multiplanet, Christiansen-multiplicity-2022IAU}.
Blindly applying a single threshold 0.99 on the posterior for two different unknown signals, one around a target star with multiple existing exoplanets and another around a target star with known FPs, is not optimal. One might want to adjust this threshold or equivalently update the score of the classifier based on the multiplicity information. The latter is the focus of this work.  

We propose to incorporate multi-planet configuration information to update the posterior probability of a given model in order to more accurately validate new exoplanets in the \kepler~\citep{borucki2010kepler} data. Our framework is agnostic to the type of the base classifier and can be applied to any model, statistical or machine learning. We show that incorporating such configuration information to an existing classifier improves its performance. By applying this method to a new version of the \ExoMiner\ model, we validate 69 new exoplanets in the \kepler\ Objects of Interest (KOI) list from the Q1-Q17 DR25 KOI catalog~\citep{Thompson_2018}. These new exoplanets are related to transit signals around target stars whose specific configuration increases the likelihood of these signals being associated with transiting exoplanets.

The new version of \ExoMiner, which we call \ExoMiner\ V1.2, addresses some of the shortcomings of the original model~\citep{Valizadegan_2022_ExoMiner}. \ExoMiner\ V1.2 receives multiple new inputs that are useful for the correct classification of FPs due to eclipsing binaries and background sources.

\section{Background}
\label{sec:background}

\subsection{Setup}
\label{sec:setup}
In this paper, we assume that there is a classifier able to provide a score for any given transit signal. We also assume that such a classifier does not have access to the system configuration of the target star; i.e., it does not use the dispositions of other transit signals detected for the target star. Existing transit signal classifiers in this category include \vespa~\citep{Morton-2016-vespa}, \Autovetter~\citep{Jenkins-Autovetter-2014IAUS,McCauliff_2015}, \Robovetter~\citep{Coughlin2017robovetter}, \AstroNet~\citep{shallue_2018}, \ExoNet~\citep{Ansdell_2018}, \GPC, \RFC, and other classifiers introduced in~\cite{Armstrong-2017-SOM}, and \ExoMiner~\citep{Valizadegan_2022_ExoMiner}. 

The objective of this work is to utilize the knowledge of the configuration of a target star in order to improve the performance of a given transit signal classifier. In our discussion, we distinguish between two sources of information: 1) the multiplicity information of the system defined by the number of Confirmed Planets (CPs), FPs, and unknown (i.e.,~unclassified) transit signals for that system and 2) the transit information of a signal. We denote the multiplicity and transit information of signal $x$ by $multi(x)$ and $transit(x)$, respectively. 

Let us denote by $s_f(x)$ the score of a classifier $f$ operating only on transit information $transit(x)$\footnote{$f$ does not have access to $multi(x)$.}. The multiplicity information $multi(x)$ is represented by a triple $N_{FPs}(x)$, $N_{CPs}(x)$, and $N_{UNKs}(x)$, which are the total number of FPs~\citep{bryson2017kepler}, the total number of CPs, and the total number of unknown objects of interest~\citep{Thompson_2018}, respectively, for the star hosting signal $x$; i.e., $multi(x)= [N_{FPs}(x), N_{CPs}(x), N_{UNKs}(x)]$. We aim to build a model $g$ that combines the multiplicity and transit information in order to build a more accurate classifier. 

In our discussion, we denote an exoplanet in the output from a model by $y=1$ and an FP by $y=0$.

\subsection{Preliminary Work}
\label{sec:preliminary_work}
Before developing the multiplicity boost framework of this work that we will discuss in Section~\ref{sec:methodology}, we report the results of applying the multiplicity boost framework of~\cite{Lissauer_2012_multiplanet, Lissauer_2014_multiplanet, Rowe_2014_multiplanet} in order to boost the performance of an existing classifier. This approach proved to be unsatisfactory, motivating the development of a data-driven approach introduced in Section~\ref{sec:methodology}. 

~\cite{Lissauer_2012_multiplanet, Lissauer_2014_multiplanet, Rowe_2014_multiplanet} introduced a framework to compute the probability of a transit signal being associated with a planet using multiplicity boost. Their framework assumes that ``false positives are randomly distributed among the targets''  and that ``there is no correlation between the probability of a target to host one or more detectable planets and to display false positives''. Their working data set was the Q1-Q8 KOI catalog for which few gold standard labels were known at that point.. Therefore, to test their model,~\cite{ Lissauer_2014_multiplanet, Rowe_2014_multiplanet} used planet candidates and identified FPs to build two data sets: one that included low-depth Eclipsing Binaries (EB) (depth $<2\%$) and another that did not. After counting the total number of targets in the sample with $i$ planet candidates for different values of $i$ ($i = 1,2,3,\ldots$), they estimated two parameters in a data driven approach: 1) the fidelity of the sample of single planet candidates, $P_1$, and 2) the number of existing FP planet candidates present among the multi-planet systems, $n_{fm}$. To this end, they estimated $P_1$ as the number of single planet candidates that are actual planets by excluding identified FPs from the population of singles, and $n_{fm}$ using the number of known observations for different scenarios (e.g., number of targets with two and three FPs, one planet + one FP, one planet + two FPs). They showed that this approach provided predictions of these different scenarios in close agreement to the observed values. 

Denoting by $p_{multi}(y=1|x)$ the probability of exoplanet using the multiplicity information alone, one can utilize the multiplication strategy introduced in~\cite{armstrong-2020-exoplanet} and adapted by~\cite{Valizadegan_2022_ExoMiner} in order to combine the probability score of the classifier $f$, i.e. $p_{f}(y=1|x)=s_f(x)$, with $p_{multi}(y=1|x)$ as follows:

\begin{equation}
\label{eq:multiplicity_information}
p_{f,multi}(y=1|x)=\frac{p_f(y=1|x)p_{multi}(y=1|x)}{\sum_y p_f(y|x) p_{multi}(y|x)}.
\end{equation}
For this to work, we need a machine classifier $f$ able to generate meaningful probability scores, i.e., $s_f(x)=p_f(y=1|x)$. 

To calculate $p_{multi}(y=1|x)$, we applied the statistical framework of~\cite{Lissauer_2012_multiplanet, Lissauer_2014_multiplanet, Rowe_2014_multiplanet} to the more recent \kepler\ Q1-Q17 DR25 KOI catalog~\citep{Thompson_2018} by performing similar stellar and KOI cuts to our initial target star population. The stellar catalog contains a total of 104,159 targets ($n_t$=104,159) that pass the following constraints: 3-hour transit duration combined differential photometric precision~\citep[CDPP, ][]{Christiansen_2012_CDPP, Jenkins2020KeplerHandbook} smaller than 160 ppm; not evolved as defined by \cite{huber2014}; not binary per the analysis of \cite{Berger_2018};  with a renormalized unit weight error (RUWE, See Section~\ref{Sec:exominer-v1.2}) less than 1.2;
and with an observation duty cycle greater than 25\%. 
For those \kepler\ Q1-Q17 DR25 KOIs that are already confirmed as planets or certified as FPs~\citep{Bryson-2015-certifiedlist}, we use their corresponding label. For those not listed as confirmed or certified as FP, we used the Exoplanet Archive disposition in the Cumulative KOI table (i.e., candidate or FP). 
 
Table~\ref{table:input_to_row} shows the input candidate abundances used to run the statistical framework~\cite[similar to Table 2 in ][]{Lissauer_2014_multiplanet}. Using ordinary least squares, we minimized the squared error between the input value for $n_{fm}$ and the predicted value based on the estimates of expected number of FPs for the different scenarios. We used as initial values for $n_{fm}$ in our optimization process the values that were estimated in~\cite{Lissauer_2014_multiplanet}. Table~\ref{table:results_rowe} reports the results of this experiment for two data sets, one that includes low-depth EBs and one that does not~\cite[similar to Table 4 in ][]{Lissauer_2014_multiplanet}. Assuming the number of observations provides a good measure of mean and the observations follow a Poisson distribution, the method in~\cite{Lissauer_2014_multiplanet} generates reasonable results (i.e.,~difference between Predicted and Observed $< 3\sigma$) for the Q1-Q17 DR25 data.

\begin{table}[!ht]
\caption{Input target counts for our experiments for the DR25 data set. $n_t$ is the number of targets from which the sample is drawn and $n_i, i\in \left\{1, \ldots , 7\right\}$ are the number of targets with exactly $i$ KOIs. With and without EBs refer to the cut for low-depth EBs as per ~\citet{Rowe_2014_multiplanet}.} 
\label{table:input_to_row}
\begin{threeparttable}
\footnotesize
\begin{tabularx}{\linewidth}{@{}Y@{}}
\begin{tabular}{c|cccccccc }
\toprule
 & $n_t$ & $n_1$ & $n_2$ & $n_3$ & $n_4$ & $n_5$ & $n_6$ & $n_7$ \\
\midrule
Without EBs & 104159  & 3215 & 383 & 127 & 47 & 17 & 4 & 1\\
With EBs & 104159  & 3714 & 406 & 128 & 47 & 17 & 3 & 2\\
\bottomrule
\end{tabular}
\end{tabularx}
\end{threeparttable}
\end{table}
 
\begin{table}[!t]
\footnotesize
\caption{Comparison between number of predicted and identified FPs in multis using the statistical framework of~\cite{Rowe_2014_multiplanet} for the DR25 data set. $n_{fm}$ is the expected total number of identified FP candidates in multis and is the sum of the rows. We followed ~\cite{Rowe_2014_multiplanet} by running two tests: one with low-depth EBs and one without. 
}
\resizebox{.93\columnwidth}{!}{
\begin{threeparttable}
\begin{tabularx}{\linewidth}{@{}Y@{}}
\begin{tabular}{c|cc|cc }
\toprule
Scenarios & \multicolumn{2}{c|}{With EBs} & \multicolumn{2}{c}{Without EBs}\\
\midrule
\empty & Predicted & Observed &  Predicted & Observed\\
\midrule
2 FPs &	41.20 &	34 & 22.74 & 17\\
3 FPs &	0.27 &	0 & 0.11 & 0\\
1 planet + 1 FP & 40.45  & 30 & 30.13 & 23\\
1 planet + 2 FPs & 0.72 & 1 & 0.39 & 1\\
$\geq 2$ planets + 1 FP & 13.29 & 7 & 9.68 & 6\\
$\geq 2$ planets + 2 FPs & 0.24 & 0 & 0.12 & 0\\
\midrule
$n_{fm}$ & 96.3 & 72 & 63.25 & 47\\
\bottomrule
\end{tabular}
\end{tabularx}
\end{threeparttable}}
\label{table:results_rowe}
\end{table}

However, the probability of observing 72 or fewer FPs for the ``with EBs'' case when the expected number is 96.3 is 0.59\%. Similarly, the probability of observing 47 or fewer FPs for the ``without EBs'' case when the expected number is 63.25 is 2.0\%. These results indicate that the model is not a particularly accurate representation of the DR25 data set.

Therefore, instead of using the multiplicity boost model from~\cite{Lissauer_2014_multiplanet, Rowe_2014_multiplanet}, we develop a fully machine learning model to boost the probability scores of a KOI using the information available for the other KOIs of the host star. A machine learning model is more flexible in that 1) it does not require the base classifier $f$ to generate probability scores; the generated scores can have any range, and 2) we do not have to use the multiplicity rule introduced in Equation~\ref{eq:multiplicity_information}. Instead, we can directly learn the mapping from inputs that consist of multiplicity information and classifier scores to the output label (exoplanet or FP). This is the focus of the following section.

\section{Multiplicity Boost}
\label{sec:multiplicity_boost}

Equation~\ref{eq:multiplicity_information} provides a way to compute the posterior probability using two initial probabilities: $p_{multi}$ and $p_f$ that are computed using two sources of information, $multi(x)$ and $transit(x)$, respectively. However, as we discussed in Section~\ref{sec:preliminary_work}, these results obtained from the multiplicity boost model of~\cite{Rowe_2014_multiplanet} is not a particularly accurate representation of the DR25 data set. Therefore, we propose to directly learn the posterior probability of a transit signal being a planet using the combination of multiplicity information $multi(x)$ and transit information $transit(x)$ in a fully data driven way. This is the focus of this section.

\subsection{Proposed Methodology}
\label{sec:methodology}

\begin{figure}[htb!]
	\centering
	\subfigure[Option 1 ]{\label{fig:multiplicity_boost_base}\includegraphics[width=\columnwidth]{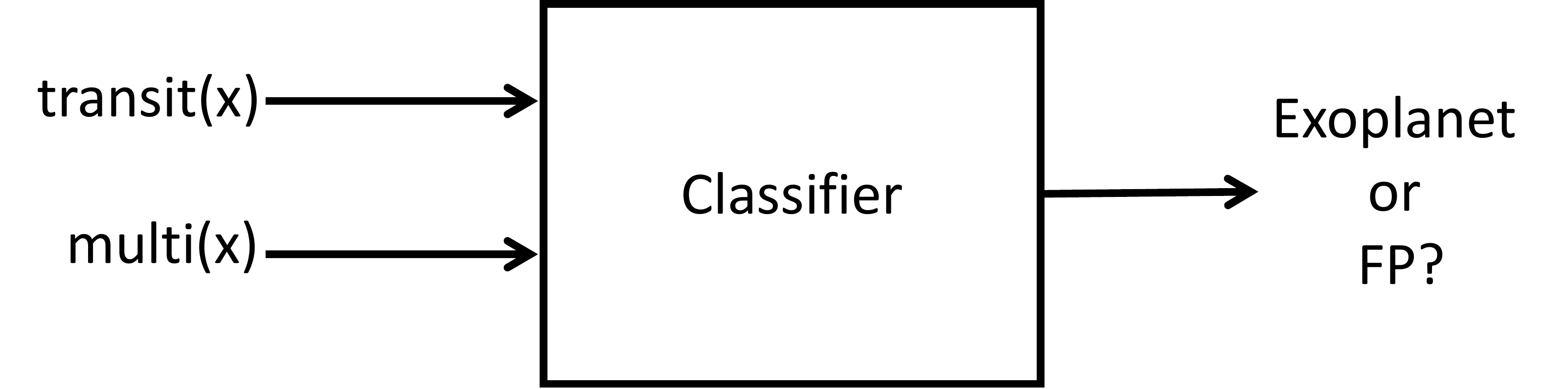}}
	\subfigure[Option 2]{\label{fig:multiplicity_boost_base_us}\includegraphics[width=\columnwidth]{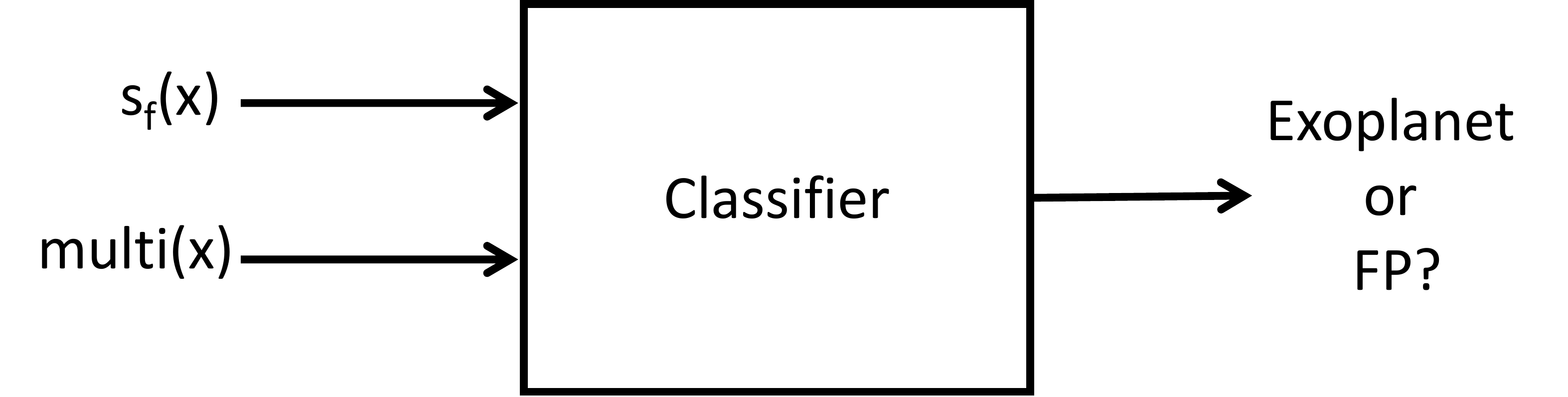}}
	\caption{Two machine learning models to use multiplicity boost information.} 
\label{fig:secondary-example}
\end{figure}

In order to directly learn the posterior probability of a transit signal being a planet using the combination of multiplicity information $multi(x)$ and transit information $transit(x)$, we propose to train a classifier that receives the multiplicity and transit information as input and generates a posterior probability, as depicted in Figure~\ref{fig:multiplicity_boost_base}. 

In this work, rather than training a single classifier to use both the transit and multiplicity information directly, we train a classifier that receives the score provided by an existing transit classifier $f$, i.e., $s_f(x)$, and multiplicity information $multi(x)$ to return the probability of exoplanet, depicted in Figure~\ref{fig:multiplicity_boost_base_us}. This approach has multiple advantages including:
\begin{itemize}
    \item Transit signal classifiers often work on Threshold Crossing Events (TCEs) but multiplicity information is only meaningful for FP and planetary candidate KOIs. By separating the model that uses transit information from the one that uses multiplicity information, we make the best use of each source of information. 
    \item This provides a clear design that can be applied on the results of any existing classifier without defining a new classifier. As an example, incorporating multiplicity information into \Robovetter\ could be very difficult. This is because \Robovetter~\citep{Coughlin_2016_robovetter} is an expert system classifier that leverages manually incorporated domain knowledge in the form of if-then conditions to check different types of diagnostic values and transit fit values in order to classify a TCE. Adapting it to accept multiplicity boost information is not straightforward. 
    \item The multiplicity information does not interact with the transit information (e.g., diagnostic metrics and plots) directly. Thus, there is no clear advantage of combining them as input of a single classifier. 
\end{itemize}

In order to combine the score of an existing classifier with the multiplicity information of the system, we build a multiplicity classifier $g$ that receives $s_f(x)$, $N_{FPs}(x)$, $N_{CPs}(x)$, and $N_{UNKs}(x)$ as inputs and generates a score $s_g(x)$. In order to train $g$, we build a training set that consists of input/output pairs in which the input is a quadruple, $[s_f(x), N_{FPs}(x), N_{CPs}(x), N_{UNKs}](x)]$, and the output is whether $x$ is a planet ($y=1$) or an FP ($y=0$). For the remaining of this paper, we remove $(x)$ and use $[s_f, N_{FPs}, N_{CPs}, N_{UNKs}]]$ to simplify the notation. 

\begin{figure}[ht]
\begin{center}
\centerline{\includegraphics[width=.5\textwidth]{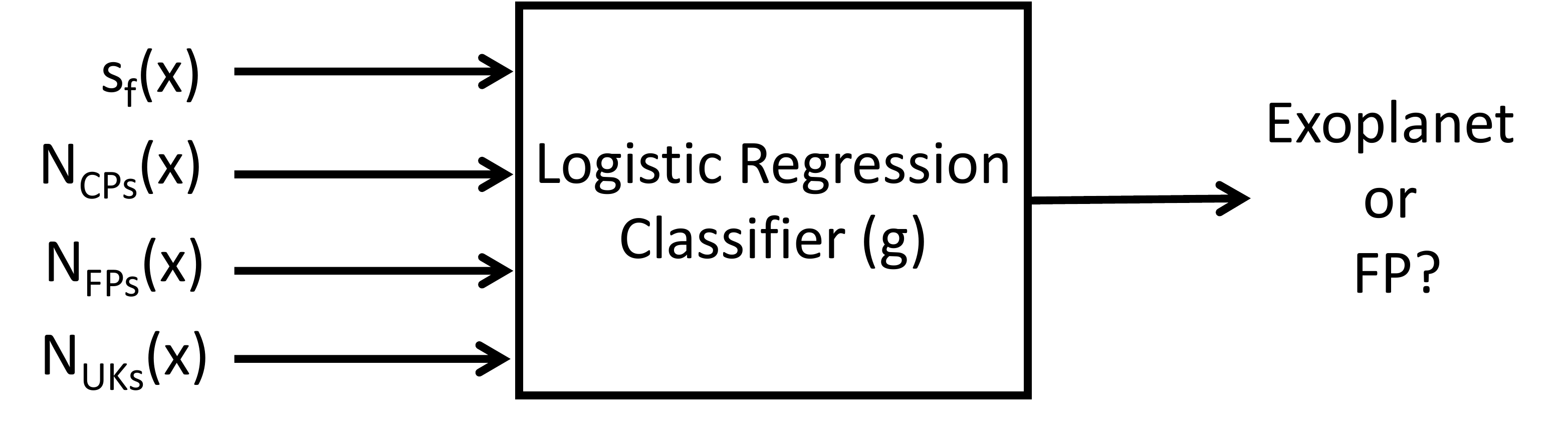}}
\caption{The machine learning model to boost the performance of a given classifier using multiplicity information.}
\label{fig:multiplicity_boost}
\end{center}
\end{figure}

We can train any machine classifier for $g$. In this work, we train a logistic regression model~\citep{cox-1958-logisticregression, Bishop:2006:PRM:1162264} for two reasons: 1) it generates meaningful probability scores that do not need calibration and can be immediately used for validation and 2) it is especially useful when input variables include discrete ordinal variables (e.g., counts). The diagram of this system is depicted in Figure~\ref{fig:multiplicity_boost}. We provide a brief review of logistic regression below.

Under rather general assumptions~\citep{Bishop:2006:PRM:1162264}, the posterior probability of class $y=1$ can be written as a logistic sigmoid function operating over the linear function of input $x$, i.e.,

\begin{equation}
\label{eq:logistic_regression}
p_g(y=1|x)=\sigma(w'x)=\frac{1}{1+\exp{(-w'x)}}.
\end{equation}
where $'$ indicates the transpose operator. In statistics, this model is called logistic regression. Unlike its name, this model is designed for classification problems~\citep{cox-1958-logisticregression}, not regression. 
Optimizing this model with regard to the linear parameter, $w$, is achieved by maximum likelihood over the observed data. Our observed data includes pairs of $<x_i,y_i>,\;i\in\left\{1, \ldots, n\right\}$, where $x_i$ is a quadruple $[s_f,\;N_{FPs},\;N_{CPs},\; N_{UNKs}]_i$ (the inputs to the model in Figure~\ref{fig:multiplicity_boost}), and $y_i$ is its label, exoplanet or FP (right side of Figure~\ref{fig:multiplicity_boost}). Denoting $p_g(y|x_i)$ by $g_i$ and having $\bm{g}=(g_1,..,g_n)$, the likelihood function can be written as:  
\begin{equation}
\label{eq:logistic_regression_likelihood}
p_g(\bm{g}|w)=\prod_{i=1}^Ng_i^{y_i}(1-g_i)^{1-y_i},
\end{equation}
which can be maximized over the observed data to learn $w$. 

For the multiplicity boost framework of this work, which has four input data elements, $w'x$ in Equation~\ref{eq:logistic_regression} can be expanded as follows:
\begin{align}
\nonumber
w'x&=w_{s_f}\times s_{f}+w_{N_{FPs}}\times N_{FPs}\\
&+w_{N_{CPs}}\times N_{CPs}+w_{N_{UNKs}}\times N_{UNKs}.
\end{align}

The weights in logistic regression have a nice interpretation. Note that:
\begin{align}
\label{eq:logodd}
\log{\biggl(\frac{p_g(y=1|x)}{1-p_g(y=1|x)}\biggr)}&=w'x.
\end{align}
Given that $w=[w_{s_f},\;w_{N_{FPs}},\;w_{N_{CPs}},\;w_{N_{UNKs}}]$ and $x=[s_f,\;N_{FPs},\;N_{CPs},\; N_{UNKs}]$, this exponential relationship provides an interpretation for the weights; when the value of an input changes, the odds will change exponentially proportional with the change. As an example, the odds are multiplied by $\exp{(w_{N_{FPs}})}$ when we increase $N_{FPs}$, the total number of FPs, by 1. After training the logistic regression and knowing the values of $w$, we know how each input affects the output.

\subsection{Constructing the Training Set}
\label{sec:construting-training-set}
To train the multiplicity boost framework described in the previous section, we need to have 1) a base classifier that generates a score for a given transit signal and 2) access to multiplicity information of the target stars in the data set. We assume a transit signal classifier is already trained and provided. Here, we discuss how to construct the multiplicity information. Assume we have access to a catalog of stars and their number of known planets, FPs, and other detected transits that are not annotated yet. Without loss of generality and to simplify the discussion, we focus on \kepler\ data to describe the details of how we construct our training set. However, the same approach can be applied to data of any other survey. 

In the \kepler\ catalog, each star is associated with multiple labeled (exoplanets and FPs) and unlabeled (unknowns) KOIs. Given that our objective is to to train the logistic regression model depicted in Figure~\ref{fig:multiplicity_boost}, we need to build input/output pairs $\big<\big[s_f, N_{FPs}, N_{CPs}, N_{UNKs}\big], y(x)\big>$ where $x$ is a transit signal and $y$ is the corresponding label (exoplanet or FP). 

A given annotated KOI provides one single input/output pair that can be used to generate more pairs in order to create a uniform data set. To this end, we construct additional input/output pairs from a single KOI by combinatorially assuming that each subset of annotated KOIs (CPs or FPs) for the target star could be unknown. To explain this, suppose there is exoplanet B on a target star with one FP, two other CPs (hence, three planets in total), and one unknown KOI. The original pair for this KOI is $\big<\big[s_f(B), 1, 2, 1\big], y=1\big>$. We build the following pairs out of this single KOI by turning each exoplanet or FP to unknown: 
\begin{enumerate}[itemsep=-1mm]
    \item Original: $\big<\big[s_f(B), 1, 2, 1\big], y=1\big>$
    \item Changing one CP to unknown: $\big<\big[s_f(B), 1, 1, 2\big], y=1\big>$
    \item Changing two CPs to unknown: $\big<\big[s_f(B), 1, 0, 3\big], y=1\big>$
    \item Changing one FP to unknown: $\big<\big[s_f(B), 0, 2, 2\big], y=1\big>$
    \item Changing one FP and one CP to unknown: $\big<\big[s_f(B), 0, 1, 3\big], y=1\big>$
    \item Changing one FP and two CPs to unknown: $\big<\big[s_f(B), 0, 0, 4\big], y=1\big>$
\end{enumerate}
For a KOI on a target star that has $N_{FPs}$ FPs and $N_{CPs}$ CPs (excluding that KOI), we build a total of $(N_{FPs}+1)\times(N_{CPs}+1)$ new instances. The idea here is that the logistic regression model in Figure~\ref{fig:multiplicity_boost} needs to learn the mapping even when a labeled KOI is assumed unknown. We aim to provide all combinations of inputs to the model so that it can generalize better to unseen situations that can have any subset of the KOIs labeled or not. 

This procedure, however, repeats the same KOIs multiple times, which leads to more emphasis on KOIs for systems with more known companion KOIs. To address this, we re-weight each generated sample by the number of examples we generated from the original KOI, i.e., $\left[(N_{FPs}+1)\times(N_{CPs}+1)\right]^{-1}$. For instance, in the above example, we re-weight each generated sample by $1/6$. Utilizing weights for samples in machine learning models, including logistic regression, is a standard practice in imbalanced cases. Even if a classifier does not accept such weights, one can simply repeat the samples in order to incorporate their appropriate weights (e.g., with probabilistic sampling with replacement).

\section{Proof of Concept}
As a proof of concept prior to using the proposed framework to validate new exoplanets, we show that the methodology introduced in Section~\ref{sec:multiplicity_boost} can improve the performance of any given classifier. 

\subsection{Base Classifiers}
\label{sec:base-classifiers}
In order to show that the methodology introduced in this work can improve the performance of any given classifier, we apply this methodology to multiple classifiers described in~\cite{Valizadegan_2022_ExoMiner} and summarized below:

\begin{enumerate}[itemsep=-1mm]
\item{\textbf{vespa:}} We use \vespa\ FP probabilities provided in the \kepler\ Q1-Q25 DR25 False Positive Probability table described in~\citet{Morton_2012_vespa, Morton-2016-vespa}. 

\item{\textbf{Robovetter:}} We use the \Robovetter\ scores ~\citep{Coughlin_2016_robovetter} for \kepler\ Q1-Q17 DR25 TCE catalog. 

\item{\textbf{AstroNet:}} In~\cite{Valizadegan_2022_ExoMiner}, the authors used the \AstroNet\ code available on GitHub~\citep{shallue_2018}, preprocessed the data, and trained the model using the same setup and DNN architecture as provided in~\citet{shallue_2018}. In this work, we use the scores reported in~\cite{Valizadegan_2022_ExoMiner}. 

\item{\textbf{ExoNet:}} The original code of \ExoNet\ is not available. We use the scores of \ExoNet\ reported in~\cite{Valizadegan_2022_ExoMiner}.

\item{\textbf{Random Forest Classifier (\RFC):}} This is one of the classifiers introduced in~\citet{armstrong-2020-exoplanet}. We use the scores provided by the authors. 

\item{\textbf{Gaussian Process Classifier (\GPC):}} This is one of the classifiers introduced in~\citet{armstrong-2020-exoplanet}. We use the scores provided by the authors, similar to \RFC\ above. 

\item{\textbf{ExoMiner:}} We use the scores reported for the \ExoMiner\ classifier in~\cite{Valizadegan_2022_ExoMiner}. 
\end{enumerate}

For each one of the aforementioned classifiers, we use two sets of classifiers' scores as input to our logistic regression model: 1) the raw output of the classifier~\citep[Table 4 in ][]{Valizadegan_2022_ExoMiner} and 2) the posterior probability of the classifier after the application of prior probability of different scenarios as discussed in~\citet{Morton-2016-vespa, armstrong-2020-exoplanet, Valizadegan_2022_ExoMiner}~\citep[see also Table 11 in][]{Valizadegan_2022_ExoMiner}. The only exception is \vespa\ for which computing the posterior probability is part of the model. Thus, we only use the posterior probability as input to the logistic regression model. 
As we will discuss in Section~\ref{sec:result-baseclassifiers}, our extensive study shows that our framework is agnostic to the source of the scores and generates similar performance boosts when applied directly to the outputs of the original classifiers.


\begin{table*}[!t]
\footnotesize
\caption{The multiplicity model introduced in this paper can be applied to any given classifier to boost its performance. The top part of this table is the result of our multiplicity boost framework when the raw classifiers' scores are used as input to the logistic regression. The bottom portion of the table shows the performance of our multiplicity boost when the posterior probabilities of the classifiers are used as input to the logistic regression. As can be seen, our multiplicity boost framework is agnostic to how the original probability scores are obtained and can improve the scores independently using the multiplicity information.}
\begin{threeparttable}
\begin{tabularx}{\linewidth}{@{}Y@{}}
\begin{tabular}{c|cccc|cccc }
\toprule
  \multicolumn{9}{c}{Raw Scores} \\
\midrule
 & \multicolumn{4}{c}{Original~\citep[Table 4, ][]{Valizadegan_2022_ExoMiner}} & \multicolumn{4}{c}{Boosted Using Our Multiplicity Approach}  \\
Model/Metric & Precision \& Recall & Accuracy & PR AUC & ROC AUC & Precision \& Recall& Accuracy & PR AUC & ROC AUC\\
\midrule
  \Robovetter & 0.951 \& 0.975 & 0.994 & 0.958 & 0.994 & 0.955 \& 0.98 & 0.995 & 0.975 & 0.999  \\

  \AstroNet & 0.861 \& 0.885 & 0.981 & 0.925 & 0.993 & 0.875 \& 0.921 & 0.984 & 0.959 & 0.996 \\

 \ExoNet & 0.925 \& 0.864 & 0.985 & 0.956 & 0.995 & 0.922 \& 0.933 & 0.989 & 0.976 & 0.998  \\
 
 \GPC & 0.921 \& 0.964 & 0.991 & 0.982 & 0.998 & 0.92 \& 0.975 & 0.992 & 0.986 & 0.998  \\
 
 \RFC & 0.929 \& 0.955 & 0.991 & 0.979 & 0.998 & 0.928 \& 0.973 & 0.992 & 0.961 & 0.998\\

 \ExoMiner  &  0.968 \& 0.974 & 0.996 & 0.995 & 1.0 &  0.971 \& 0.977 & 0.996 & 0.993 & 1.0 \\
\midrule
  \multicolumn{9}{c}{Posterior Probabilities} \\
\midrule
 & \multicolumn{4}{c}{Original~\citep[Table 11, ][]{Valizadegan_2022_ExoMiner}} & \multicolumn{4}{c}{Boosted Using Our Multiplicity Approach}  \\
Model/Metric & Precision \& Recall & Accuracy & PR AUC & ROC AUC & Precision \& Recall& Accuracy & PR AUC & ROC AUC\\
\midrule
  \vespa &  0.666 \& 0.968 & 0.801 & 0.865 & 0.928 & 0.749 \& 0.973 & 0.865 & 0.895 & 0.943  \\

  \Robovetter & 0.947 \& 0.979 & 0.994 & 0.964 & 0.997 & 0.954 \& 0.981 & 0.995 & 0.975 & 0.999  \\

  \AstroNet & 0.920 \& 0.920 & 0.988 & 0.968 & 0.996 & 0.921 \& 0.955 & 0.990 & 0.980 & 0.998 \\

 \ExoNet & 0.933 \& 0.902 & 0.988 & 0.97 & 0.997 & 0.933 \& 0.953 & 0.991 & 0.983 & 0.998  \\
 
 \GPC & 0.910 \& 0.977 & 0.991 & 0.987 & 0.998 & 0.911 \& 0.983 & 0.992 & 0.986 & 0.998  \\
 
 \RFC & 0.942 \& 0.964 & 0.993 & 0.983 & 0.999 & 0.942 \& 0.978 & 0.994 & 0.965 & 0.999\\

 \ExoMiner  &  0.971 \& 0.978 & 0.996 & 0.996 & 0.999 & 0.974 \& 0.983 & 0.997 & 0.993 & 1.000 \\
\bottomrule
\end{tabular}
\end{tabularx}
\end{threeparttable}
\label{table:performance_boost}
\end{table*}

\subsection{Evaluation Metrics}
We compare the performance using the following metrics:
\begin{itemize}
\item{\textbf{Accuracy:}} This is the fraction of correctly classified cases. When the data set is imbalanced in terms of the percentage of positive examples, this metric is not particularly informative. To see this, note that a classifier that classifies all transits as FPs has an accuracy of 90\% if 90\% of examples are FPs. However, accuracy provides some insights when studied in conjunction with other metrics.

\item{\textbf{Precision:}} Also called positive predictive value, this is the fraction of transit signals classified as planets that are indeed true planets; i.e.,
\begin{equation}
\nonumber
Precision=\frac{true\;positives}{true\;positives+false\;positives}
\end{equation}

\item{\textbf{Recall:}} Also called the true positive rate, this is the fraction of planets correctly classified as such; i.e.,
\begin{equation}
\nonumber
Recall=\frac{true\;positives}{true\;positives+false\;negatives}
\end{equation}

\item{\textbf{Precision-Recall (PR) curve:}} The PR curve summarizes the trade-off between precision and recall by varying the threshold used to convert the classifier's score into a label. PR area under the curve (AUC) is the total area under the PR curve. An ideal classifier would have an PR AUC of 1.

\item{\textbf{Receiver Operating Characteristic (ROC) curve:}} The ROC curve summarizes the trade-off between the true positive rate (recall) and false positive rate (fall-out) when varying the threshold used to convert the classifier score into a label. ROC AUC is the total area under the ROC. 

\end{itemize}

\subsection{Performance Boost of Existing Classifiers}
\label{sec:result-baseclassifiers}
To test and compare the performance of different base classifiers (with and without priors), we use the same labels and 10-fold cross-validation scheme reported in~\cite{Valizadegan_2022_ExoMiner}. The performance of these classifiers before and after the application of our multiplicity boost methodology is reported in Table~\ref{table:performance_boost}.  As can be seen from the results, the multiplicity information improves the performance of all classifiers across all evaluation metrics, with the exception of PR AUC for some models. The reason why the multiplicity boost information reduces the performance in terms of PR AUC in some cases is because logistic regression was not designed to optimize PR AUC directly. Overall, the proposed method is effective and independent of the type of the employed base classifier.

\begin{table*}[!t]
\footnotesize
\caption{Logistic regression weights for different classifiers. }
\begin{threeparttable}
\begin{tabularx}{\linewidth}{@{}Y@{}}
\begin{tabular}{c|cccc|cccc}
\toprule
& \multicolumn{4}{c}{Original~\citep[Table 4, ][]{Valizadegan_2022_ExoMiner}}  & \multicolumn{4}{c}{Posterior Probabilities~\citep[Table 11, ][]{Valizadegan_2022_ExoMiner}} \\
\midrule
Model/Parameter & $w_{s_f}$ & $w_{N_{FPs}}$ & $w_{N_{CPs}}$ & $w_{N_{UNKs}}$ & $w_{s_f}$ & $w_{N_{FPs}}$ & $w_{N_{CPs}}$ & $w_{N_{UNKs}}$\\
\midrule
\vespa &  &  & & &  $6.20\pm 0.05$ & $-1.93\pm 0.05$ & $4.63\pm0.07$ & $2.60\pm0.02$  \\
\Robovetter & $7.79\pm 0.03$ & $-2.49\pm0.07$ & $3.27\pm0.06$ & $2.06\pm0.03$  & $7.90\pm 0.03$ & $-2.24\pm 0.09$ & $3.24\pm0.06$ & $2.07\pm0.04$\\

\AstroNet & $6.54\pm 0.02$ & $-2.77\pm0.06$ & $3.54\pm0.04$ & $1.93\pm0.02$  & $8.02\pm 0.03$ & $-2.02\pm 0.08$ & $3.33\pm0.04$ & $1.96\pm0.02$ \\

\ExoNet & $8.98\pm 0.03$ & $-2.51\pm0.09$ & $3.25\pm0.05$ & $1.73\pm0.02$  & $9.64\pm 0.03$ & $-2.73\pm 0.10$ & $3.06\pm0.05$ & $1.78\pm0.02$ \\
 
\GPC & $10.81\pm 0.04$ & $-1.63\pm0.03$ & $1.76\pm0.05$ & $1.00\pm0.03$   & $10.46\pm 0.03$ & $-2.16\pm 0.04$ & $1.68\pm0.05$ & $1.03\pm0.03$  \\
 
\RFC & $9.34\pm 0.03$ & $-2.36\pm 0.16$ & $2.39\pm0.04$ & $1.47\pm0.02$ & $9.76\pm 0.04$ & $-3.00\pm 0.11$ & $2.13\pm0.04$ & $1.36\pm0.03$\\

\ExoMiner  &  $9.04\pm 0.03$ & $-1.10\pm0.10$ & $2.10\pm0.05$ & $1.46\pm0.02$  &  $9.35\pm 0.03$ & $-1.46\pm0.11$ & $2.04\pm0.06$ & $1.54\pm0.03$ \\
\bottomrule
\end{tabular}
\end{tabularx}
\end{threeparttable}
\label{table:classifier_weights}
\end{table*}

\subsection{Model Interpretation} 

To study the behavior of our machine learning multiplicity boost model, we report the weight values of the logistic regression trained for different classifiers in Table~\ref{table:classifier_weights}. As mentioned in Section~\ref{sec:methodology}, the weight directly demonstrates the effect of the corresponding input parameter on the odds of the event, i.e., the transiting signal is an exoplanet. The weight for the number of FPs, i.e., $N_{FPs}$ is negative, while the weight for all other input parameters, i.e., $S_f$, $N_{CPs}$, and $N_{UNKs}$, are positive for all base classifiers. This shows that the model learned from the data to decrease the odds for systems that already have FPs and increase the odds for systems that have existing CPs or unknown KOIs. The amount of change in the odds, compared to the original odds generated by the original classifier, depends on multiple factors, which include: 1) the accuracy of the base classifier, 2) the precision/recall trade-off of the base classifier, and 3) the regions where the classifier makes more mistakes (e.g., single or multi-planet systems). In general, however, the weight for the original classifier, $s_f$, is higher for more accurate classifiers, as can be seen by comparing the numbers in Table~\ref{table:performance_boost} and Table~\ref{table:classifier_weights}. Moreover, the weights for multiplicity information, i.e., $N_{FPs}$, $N_{CPs}$, and $N_{UNKs}$, are generally smaller than the weight for $s_f$. This is because 1) the base classifier uses transit data, which is more informative than the multiplicity data and 2) the range of the base classifier score is only between 0 and 1, smaller than that of the other input parameters.  

\begin{figure}[ht]
\begin{center}
\centerline{\includegraphics[width=1.1\columnwidth]{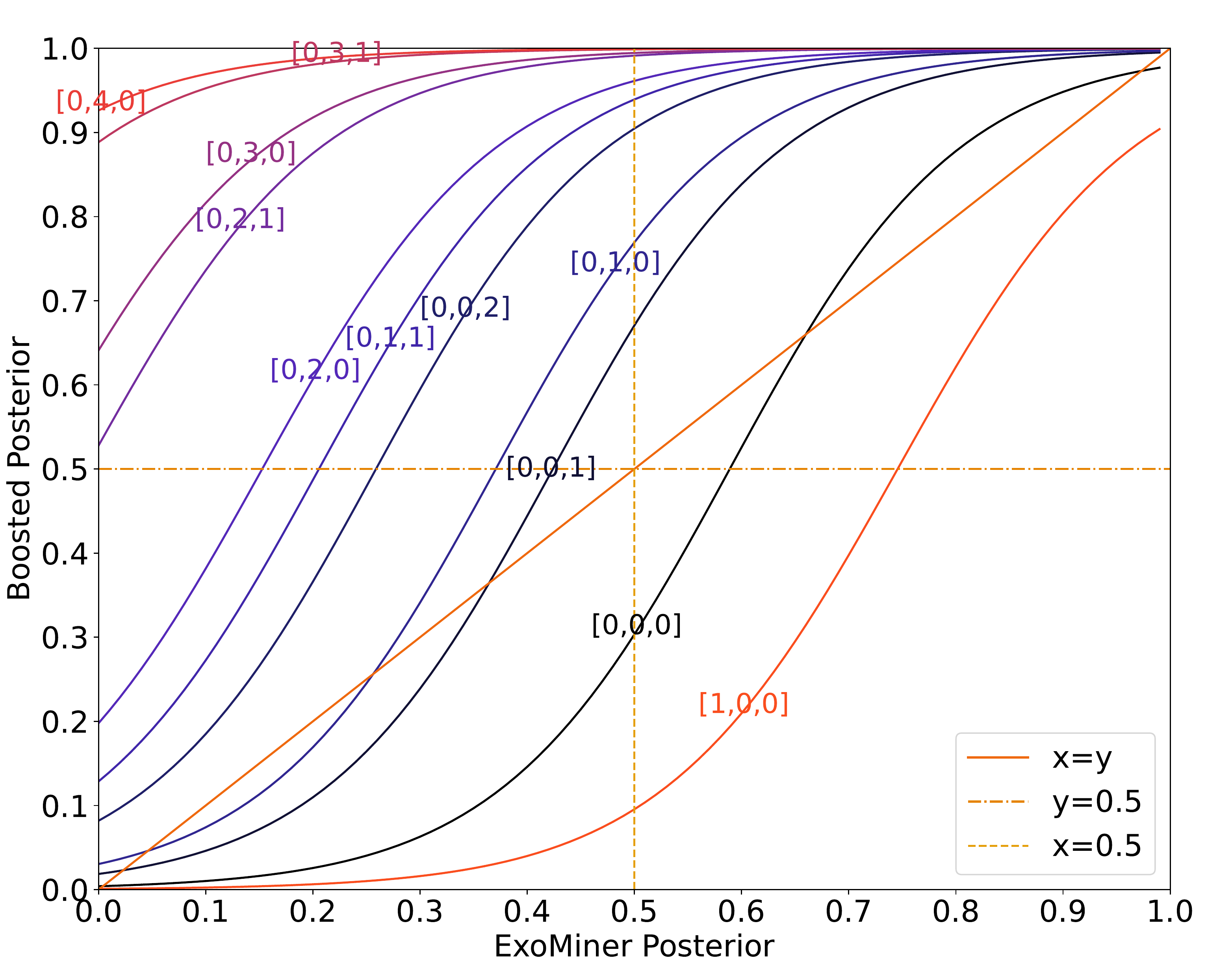}}
\caption{The mapping between \ExoMiner\ scores and the boosted scores using the logistic regression classifier. Each colored curve represents the mapping for a different multiplicity scenario $\left[N_{FPs}(x),\; N_{CPs}(x),\; N_{UNKs}(x)\right]$. Scenarios are only plotted if they have more than ten counts in the \kepler\ Q1-Q17 DR25 KOI catalog. The larger the number of existing CPs ($N_{CPs}(x)$), the more the logistic regression classifier favors classification as an exoplanet.} 
\label{fig:multiplicity_curve}
\end{center}
\end{figure}

By studying the values of individual weights, we can also better understand the importance of different input parameters for the multiplicity model. For example, the weight for $N_{CPs}$ when \vespa\ is used as the base classifier, is 4.629 which means that odds of exoplanet will increase by a factor of $\exp{(4.629)}=102.41$ for any addition of a CP to the system. This number is  $\exp{(2.043)}=7.71$ for posterior from \ExoMiner. The collective behavior can be summarized in mapping plots, where the x-axis is the score of the original classifier, the y-axis is the score generated by the multiplicity boost framework, and the different curves show the mapping for specific multiplicity scenarios $[N_{FPs}$, $N_{CPs}$, $N_{UNKs}]$. Figure~\ref{fig:multiplicity_curve} shows such a mapping for the posterior probability of \ExoMiner. For each multiplicity scenario $\left[N_{FPs}(x),\; N_{CPs}(x),\; N_{UNKs}(x)\right]$, one solid line represents the mapping for that scenario. We note a few observations here: 
\begin{enumerate}[itemsep=-1mm]
\item When the total number of existing exoplanets or unknown KOIs increases for a star, the chances that a new KOI is an exoplanet increases; In the extreme case when there are four existing exoplanets and no FPs, the multiplicity boost predicts with high confidence that the next KOI is an exoplanet independent of the score of the original classifier. As we will discuss in Section~\ref{sec:caveats}, this could go wrong for very rare situations when a background object is detected on a multi-planet system. We will introduce mitigating solutions in Section~\ref{sec:vetoes} to avoid validating such cases. 
\item When the total number of FPs increases, the probability that a new KOI is an exoplanet decreases. However, the prediction of the model for systems with more than two existing FPs should be taken with caution as there are no training data with more than three FPs and the multiplicity classifier extrapolates in that region. 
\item When the KOI being classified is the only KOI for a target star (blue line for $[0,0,0]$), the multiplicity boost framework is conservative and lowers the scores when the original scores are less than 0.57. This reflects the larger abundance of single FP KOIs. In the \kepler\ catalog, there are only 1315 exoplanets out of 4721 single KOI stars. Also note that our multiplicity boost framework does not generate scores larger than 0.99 for any single KOI, which implies that it does not validate single KOIs. Basically, no existing exoplanets in single-planet systems can be validated using this framework. This is an acceptable property because this framework is built to validate exoplanets for multi-planet systems.
\end{enumerate}

Given that this is a multiplicity boost framework, one might expect to see that the mapping for Scenario $[0,0,0]$ lies on the diagonal. This is not the case because 1) the model learns a mapping between input quadruple $\left[s_f,\; N_{FPs},\; N_{CPs},\; N_{UNKs}\right]$ and the binary output variable (planet or FP) to perform well for all scenarios; given that the model is more confident for KOIs in multi-planet systems, it adjusts its confidence of the cases for single KOIs, respectively and 2) the specific shape of the mapping is dictated by the logistic regression assumption that the log-odds for the signal being an exoplanet is a linear combination of the input variables $\left[s_f,\; N_{FPs},\; N_{CPs},\; N_{UNKs}\right]$. The combination of the aforementioned two properties leads to the specific form of the mapping curves, including one for Scenario $[0,0,0]$.

\begin{figure*}[ht]
\begin{center}
\centerline{\includegraphics[width=\textwidth]{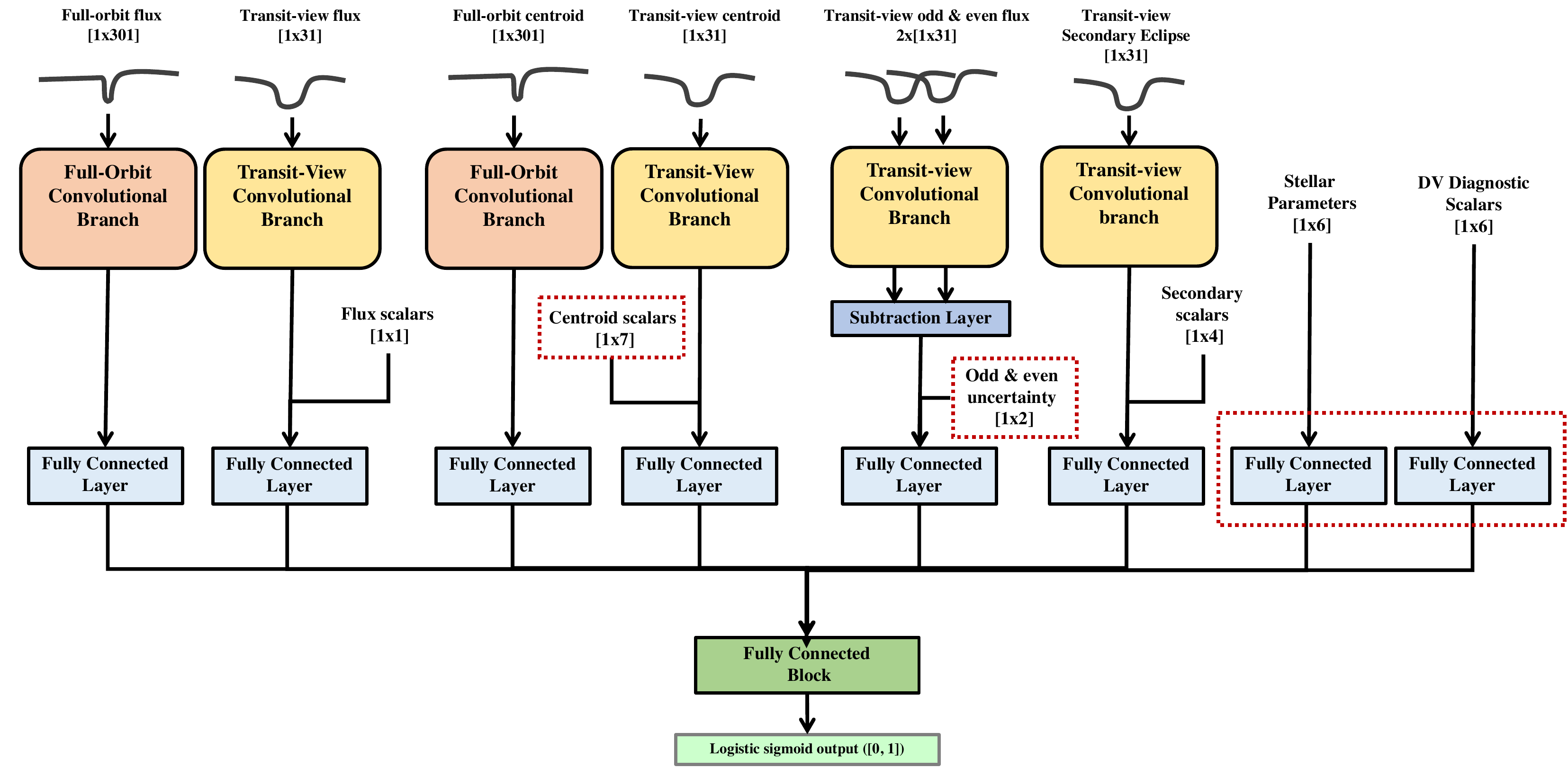}}
\caption{An adapted version of the classifier in~\cite{Valizadegan_2022_ExoMiner}. The red dotted rectangles show the differences between \ExoMiner\ and \ExoMiner\ V1.2. }
\label{fig:ExoMinerV1_2}
\end{center}
\end{figure*}

\section{ExoMiner V1.2}
\label{Sec:exominer-v1.2}
In this section, we introduce an improved version of the \ExoMiner\ classifier described in~\cite{Valizadegan_2022_ExoMiner}. This new model, which we call \ExoMiner\ V1.2, is depicted in Figure~\ref{fig:ExoMinerV1_2}. Compared to the original \ExoMiner, the new version includes the following changes itemized by the branch to which they apply:
\begin{itemize}[ leftmargin=4mm]
    \item{Transit-View Centroid Branch:} We added two new scalar values to the transit-view centroid branch: 1) target star magnitude, $K_p$, required to identify saturated stars for which the centroid diagnostic test is invalid. Instead of using the original value, we converted it to a binary value to represent whether a target star is saturated or not based on a conservative threshold $thr_{K_p}=12$, and 2) Renormalized Unit Weight Error~\cite[RUWE, ][]{Lindegren-2021-GAIA} that gives information about the quality of the transit source, i.e., larger values can indicate that the source is not a single star, or that we are in the presence of significantly crowded fields. This leads to an array of $1\times7$ centroid scalars in this branch. The other 5 features already used in the original \ExoMiner\ model are the flux-weighted centroid motion detection statistic, centroid offset to the target star according to the Kepler Input Catalog~\citep{Brown-2011-KIC}, centroid offset to the out-of-transit (OOT) centroid position, and finally the respective uncertainties for these two centroid offset estimates.
    \item{Transit-View Odd \& Even Branch:} We added one OOT variability scalar for each view (odd and even). This is related to a problem identified in the original \ExoMiner\ model that prevented it to effectively learn from the odd and even data. OOT variability scalar describes the variability in the odd/even view and is computed as the median absolute deviation (as a robust estimate of the standard deviation) for the out-of-transit cadences in the odd/even phase folded time series normalized by the square root of the number of cadences; a larger standard deviation or a lower number of cadences make this feature larger, providing information regarding the general variability of these two views.
    \item We updated the architecture of the model in~\cite{Valizadegan_2022_ExoMiner} by adding fully connected (FC) layers at the end of the stellar parameters and data validation (DV) diagnostic branches to make sure the contribution of each branch is learned through training instead of being dictated by the architecture. Similar to the other convolutional branches, these FC layers have 4 units each.
\end{itemize}
The above changes are highlighted by the red dotted rectangles in Figure~\ref{fig:ExoMinerV1_2}.

To train \ExoMiner\ V1.2, we build a data set similar to~\cite{Valizadegan_2022_ExoMiner} using the \kepler\ Q1-Q17 DR25 TCE table~\citep{twicken2016detection} and the most up-to-date planet catalog as follows: first, we removed all 1498 rogue TCEs from the list. Rogue TCEs are three-transit TCEs that were generated by a bug in the \kepler\ pipeline~\citep{Coughlin_2016_robovetter}. For the planet category, we used the TCEs that are listed as CPs in the Cumulative KOI catalog. 
For the FP class, we used: 1) TCEs in the Q1-Q17 DR25 list that are Certified FPs (CFPs) or certified false alarms (CFAs) in the \kepler\ Certified False Positive table ~\citep{Bryson-2015-certifiedlist} and 2) TCEs vetted as Non-Transiting Phenomena (NTP) by \Robovetter\ (any TCE from Q1-Q17 DR25 TCE catalog that is not in the Cumulative KOI catalog). This resulted in a total of 30957 TCEs that consisted of 2643 CPs and 28314 FPs. Note that this data set is almost identical to the one used in~\cite{Valizadegan_2022_ExoMiner}; there are only more exoplanets in this data set as the result of recent validations~\citep{armstrong-2020-exoplanet, Valizadegan_2022_ExoMiner}. 

To study the behavior of different models when the training/test split changes, we perform a 10-fold cross-validation (CV), i.e., we split the data into 10 folds, each time we take one fold for test and the other 9 folds for training/validation (8 folds for training and one fold for validation). Also, similar to~\cite{Valizadegan_2022_ExoMiner}, we split TCEs by their respective target stars instead of by TCEs to remove dependency between training and test sets.

\begin{table}[htb!]
 \centering
\caption{Multiplicity scenarios with their total number of exoplanets and FP counts. A scenario is represented by a tuple $\big[N_{FPs}(x),\; N_{CPs}(x),\; N_{UNKs}(x)\big]$ accounting for all the KOIs in the system other than the CP or CFP under consideration.} 
\label{table:scenarios}
\resizebox{\columnwidth}{!}{
\begin{threeparttable}
\begin{tabularx}{\linewidth}{@{}Y@{}}
\begin{tabular}{cccc}
\toprule
Scenarios & Confirmed Exoplanets & Certified FPs & Total \\
\midrule
$[3, 0, 0]$ & 0 & 4 & 4 \\
$[1, 0, 0]$ & 6 & 90 & 96 \\
$[1, 0, 1]$ & 0 & 2 & 2 \\
$[1, 0, 2]$ & 1 & 0 & 1 \\
$[1, 1, 0]$ & 8 & 0 & 8 \\
$[1, 2, 2]$ & 3 & 0 & 3 \\
$[1, 4, 0]$ & 5 & 0 & 5 \\
\midrule
$[0, 0, 0]$ & 1315 & 3406 & 4721 \\
\midrule
$[0, 0, 1]$ & 121 & 23 & 144 \\
$[0, 0, 2]$ & 23 & 0 & 23 \\
$[0, 0, 3]$ & 1 & 0 & 1 \\
$[0, 1, 0]$ & 510 & 6 & 516 \\
$[0, 1, 1]$ & 68 & 0 & 68 \\
$[0, 1, 2]$ & 6 & 1 & 7 \\
$[0, 1, 3]$ & 2 & 0 & 2 \\
$[0, 2, 0]$ & 294 & 4 & 298 \\
$[0, 2, 1]$ & 30 & 0 & 30 \\
$[0, 2, 2]$ & 6 & 0 & 6 \\
$[0, 3, 0]$ & 136 & 0 & 136 \\
$[0, 3, 1]$ & 20 & 0 & 20 \\
$[0, 3, 2]$ & 0 & 1 & 1 \\
$[0, 4, 0]$ & 75 & 0 & 75 \\
$[0, 5, 0]$ & 6 & 1 & 7 \\
$[0, 6, 0]$ & 7 & 0 & 7 \\
\midrule
Total & 2643 & 3538 & 6181 \\
\bottomrule
\end{tabular}
\end{tabularx}
\end{threeparttable}
}
\end{table}

\subsection{Data Set for Multiplicity Model}
\label{sec:dataset}
Out of 30957 Q1-Q17 DR25 TCEs obtained to train \ExoMiner\ V1.2, 8054 TCEs are associated with KOIs in the Cumulative KOI table~\citep{Thompson_2018}. To build the training set for the multiplicity boost, we used TCEs associated with CP or CFP\footnote{Note that we did not include CFAs in our analysis because multiplicity boost is based on the statistics of astrophysical FPs and planet multiplicity, whereas FAs do not follow these statistics.} KOIs, which resulted in a total of 2643 planets and 3538 FPs. For each target, we counted the total number of KOIs, number of planets, FPs, and remaining unclassified (UNK) KOIs. This led to a total of 6181 KOIs with their scenarios and counts presented in Table~\ref{table:scenarios}. We split this table into three sections: 1) the top section that represents scenarios that have at least one FP, 2) the middle section that represents the single planet systems, and 3) the bottom section that represents scenarios that do not have any FP. We would like to emphasize that the numbers of FPs, planets, and unknowns for each KOI include only the other KOIs in the given system. Before providing some insights and in order to better explain these numbers, we discuss some specific interesting scenarios below that have only a few FP and CP counts. Note that each scenario is represented by $\left[N_{FPs},\; N_{CPs},\; N_{UNKs}\right]$.
\begin{enumerate}[itemsep=-1mm]
    \item{Scenario [0, 0, 3]:} There is only one system, K01082, with 4 KOIs of which three are unknown. The known KOI of this system, K01082.03, is a CP. This generates one count of CP for Scenario [0, 0, 3].
    \item{Scenario [1,2,2] and [0,3,2]:} The system that generates these scenarios is K02433. There are a total of 6 KOIs for this system in DR25, which includes 3 CPs, one CFP, and 2 unknown KOIs. This results in three CPs for Scenario [1,2,2] and one FP for Scenario [0,3,2]. 
    \item{Scenario [1,4,0]:} K00082 has 5 CPs and one CFP, which results in 5 CPs for this scenario and one FP for Scenario [0, 5, 0], discussed below.  
    \item{Scenario [0, 5, 0]:} There are 6 exoplanets for K00157 that results in 6 CP for scenario [0, 5, 0]. There is also K00082 that has 5 exoplanets and 1 FP, resulting in a single FP for [0, 5, 0]. 
\end{enumerate}

In general the statistics provided in Table~\ref{table:scenarios} confirm that when there is at least one FP, it is highly likely that the unknown KOI (the KOI in question) is also an FP. When there is no FP, it is highly likely that the unknown KOI is a planet. We provide some insights regarding multiplicity scenarios with at least one CP or unknown KOI but no FPs (i.e., $[x, y, z],\; x=0,\; y>0 \;or\; z>0$), and those that have at least 1 FP count (Third column in Table~\ref{table:scenarios}):
\begin{enumerate}[itemsep=-1mm]
\item {Scenario [0, 0, 1]:} This scenario has 121 CP and 23 FP counts (1st row of the 3rd section of Table~\ref{table:scenarios}), the most among scenarios with at least one CP or one unknown KOI. The list of the KOIs for this scenario is provided in Table~\ref{table:scenario_0_0_1}. Interestingly, \ExoMiner\ V1.2 classifies all the FP KOIs correctly and gives very low scores to most of them. None of them get close to the validation threshold of 0.99 after the application of the multiplicity boost. 
\item {Scenario [0, 1, 0]:} There are 6 FPs for this scenario: K00199.02, K01944.02, K02362.01, K03685.01, K03936.01, and K03954.01. These KOIs have at least one \Robovetter\ FP flag on in the Cumulative KOI table indicating that they failed one or more FP diagnostic tests. Our \ExoMiner\ V1.2 was able to correctly classify these cases with high confidence with scores 2.61E-10, 0.001, 0.407, 0.009,
1.88E-11, 1.03E-05, respectively. The scores for these KOIs get boosted to 0.03, 0.030, 0.752, 0.040, 0.027, 0.028, respectively. None of them gets close to the validation threshold 0.99.
\item {Scenario [0, 1, 2]:} The only FP for this scenario is K03741.01 with the ``Stellar Eclipse Flag'' on in the Cumulative KOI table. \ExoMiner\ V1.2 gives a very low score of 1.10E-05 to this KOI indicating that our classifier correctly classified this KOI with high confidence. The multiplicity boost framework boosts its score to 0.381. 
\item {Scenario [0, 2, 0]:} There are four FPs for this scenario: K01196.01, K01806.01, K00672.03, K01792.02. All four KOIs are flagged by at least one FP indicator in the Cumulative KOI table. \ExoMiner\ V1.2 was also able to successfully capture the FP pattern in the data classifying them with low scores of 0.072, 4.69E-04, 2.76E-06, 2.02E-4, respectively. The multiplicity boost framework boosts their scores to 0.201, 0.207, 0.182, 0.136, respectively, all well below the validation threshold. 
\item {Scenario [0, 3, 2]:} The only KOI for this scenario is K02433.05, which is an FP. This KOI is flagged by two FP flags in the Cumulative KOI table and scored 1.55E-05 by \ExoMiner\ V1.2. Given that this KOI is around a system with three CPs and two other unknown KOIs, its score gets boosted to 0.987 by the multiplicity boost framework, which is very close to the validation threshold. 
\item {Scenario [0, 5, 0]:} Note that the only FP KOI for scenario [0, 5, 0] is the 6th KOI of K00082 with five existing exoplanets. As mentioned in~\cite{Valizadegan_2022_ExoMiner}, this is incorrectly certified as FP in the Certified False Positive table. \ExoMiner\ V1.2 gives a very high score of 0.997 to this KOI which is boosted to almost 1.0 by the multiplicity boost framework. 
\end{enumerate}

Note that we did not correct this label noise in the data as there are potentially other sources of label noise. As a matter of the fact we are aware of five CPs that have been demoted to FP\footnote{https://exoplanetarchive.ipac.caltech.edu/docs/removed\_targets.html}. These are Kepler-486 b~\cite{Diaz-2014-Kepler-486b}, Kepler-492 b~\citep{Diaz-2013-Kepler-492b}, Kepler-699 b~\citep{Niraula_2022_demotion},  Kepler-840~\citep{Niraula_2022_demotion}, Kepler-854 b~\citep{Niraula_2022_demotion}. All these are single planet systems which do not affect the learning process\footnote{They have a negligible effect on the relative ratio of FPs for Scenario [0,0,0]}. However, we would like to emphasize that  machine learning models are generally robust to a significant amount of label noise. A machine learning model does not rely on the statistics for each scenario individually but rather aims to find a mapping that optimizes its objective function, in this case Equation~\ref{eq:logistic_regression_likelihood}.

From the 6181 initial scenario sets reported in Table~\ref{table:scenarios}, we generated 8458 new scenarios and their weights using the procedure described in Section~\ref{sec:construting-training-set}. 

\begin{table*}[!t]
\caption{Information related to FP cases for Scenario [0,0,1].}
\resizebox{1\textwidth}{!}{
\begin{threeparttable}
\begin{tabularx}{\linewidth}{@{}Y@{}}
\begin{tabular}{c|cccc|cc}
\toprule
 & \multicolumn{4}{c}{KOI flags} &  \multicolumn{2}{c}{Scores}\\
 \midrule
KOI/Data & Not Transit-Like & Stellar Eclipse & Centroid Offset & Ephemeris Match & \ExoMiner\ V1.2 Posterior & Multiplicity  \\
\midrule
K06137.02 & 0 &  0 &  1 &  0 &      0.497 &      0.732 \\
K00126.01 & 0 &  1 &  0 &  0 &      0.480 &      0.699 \\
K01232.01 & 0 &  1 &  0 &  0 &      0.143 &      0.067 \\
K05449.01 & 0 &  1 &  0 &  0 &      0.132 &      0.064 \\
K04388.01 & 0 &  1 &  0 &  0 &      0.113 &      0.056 \\
K03641.01 & 0 &  1 &  0 &  0 &      0.012 &      0.024 \\
K07124.01 & 0 &  1 &  0 &  0 &      0.003 &      0.018 \\
K07544.02 & 0 &  1 &  0 &  0 &      0.003 &      0.022 \\
K02882.02 & 0 &  1 &  0 &  0 &      0.000 &      0.022 \\
K00379.01 & 0 &  0 &  1 &  0 &      0.000 &      0.021 \\
K06464.01 & 0 &  1 &  0 &  0 &      0.000 &      0.020 \\
K01957.02 & 0 &  1 &  0 &  0 &      0.000 &      0.018 \\
K02159.02 & 0 &  0 &  1 &  0 &      0.000 &      0.018 \\
K06751.02 & 0 &  1 &  1 &  0 &      0.000 &      0.017 \\
K03087.01 & 0 &  0 &  1 &  0 &      0.000 &      0.018 \\
K02184.01 & 0 &  0 &  1 &  0 &      0.000 &      0.020 \\
K02050.01 & 0 &  1 &  1 &  1 &      0.000 &      0.017 \\
K02404.02 & 1 &  0 &  1 &  1 &      0.000 &      0.017 \\
K04323.02 & 1 &  0 &  1 &  1 &      0.000 &      0.022 \\
K01562.01 & 0 &  0 &  0 &  1 &      0.000 &      0.022 \\
K01731.02 & 0 &  0 &  1 &  0 &      0.000 &      0.019 \\
K03230.02 & 1 &  0 &  0 &  1 &      0.000 &      0.019 \\
K04881.02 & 0 &  0 &  0 &  1 &      0.000 &      0.019 \\
\bottomrule
\end{tabular}
\end{tabularx}
\end{threeparttable}
}\label{table:scenario_0_0_1}
\end{table*}

\begin{table*}[!t]
\caption{Performance of \ExoMiner\ V1.2 and its performance boost using multiplicity framework.}
\begin{threeparttable}
\begin{tabularx}{\linewidth}{@{}Y@{}}
\begin{tabular}{c|cccc|cccc }
\toprule
 & \multicolumn{4}{c}{Original} & \multicolumn{4}{c}{Boosted Using Our Multiplicity Approach}  \\
Model/Metric & Precision \& Recall & Accuracy & PR AUC & ROC AUC & Precision \& Recall& Accuracy & PR AUC & ROC AUC\\
\midrule
 \ExoMiner V1.2  &  0.978 \& 0.980 & 0.996 & 0.997 & 0.999 & 0.979 \& 0.984 & 0.997 & 0.995 & 1.000\\
\bottomrule
\end{tabular}
\end{tabularx}
\end{threeparttable}
\label{table:performance_boost_new}
\end{table*}

\begin{table*}[!t]
\footnotesize
\caption{Logistic regression weights for \ExoMiner\ V1.2.}
\begin{threeparttable}
\begin{tabularx}{\linewidth}{@{}Y@{}}
\begin{tabular}{c|cccc}
\toprule
Model/Parameter & $w_{s_f}$ & $w_{N_{FPs}}$ & $w_{N_{CPs}}$ & $w_{N_{UNKs}}$\\
\midrule
\ExoMiner\ V1.2  &  $9.728\pm 0.03$ & $-1.71\pm0.10$ & $2.044\pm0.07$ & $1.596\pm0.03$ \\
\bottomrule
\end{tabular}
\end{tabularx}
\end{threeparttable}
\label{table:classifier_weights_exominer_v1_2}
\end{table*}

\subsection{Performance Study}
\label{sec:performance}
Table~\ref{table:performance_boost_new} summarizes the performance result of applying \ExoMiner\ V1.2 posterior to the most recent data set discussed in the previous subsection. It also reports the results after application of the multiplicity boost framework proposed in this work. As can be seen, our multiplicity boost framework improves the performance of this new model as well. We also report in Table~\ref{table:classifier_weights_exominer_v1_2} the weights learned by the multiplicity boost framework for each input parameter. Compared to the original classifier (Table~\ref{table:classifier_weights}), \ExoMiner\ V1.2 scores obtains higher weight as it is more accurate.

\begingroup
\setlength{\tabcolsep}{2.5pt} 
\begin{table}[htb]
 \centering
\caption{The original scores of \ExoMiner\ V1.2 and the boosted scores after the application of multiplicity boost framework on 6181 known KOIs. This table describes the available columns. The full table is available online.}
\label{table:classification_results}
\begin{threeparttable}
\begin{tabularx}{\linewidth}{@{}Y@{}}
\begin{tabular}{ll}
\toprule
Column & Description \\
\midrule
KIC & KIC ID\\
TCE & TCE planet number\\
KOI name & KOI name\\
Period (days) & TCE period\\
Radius (Re) & planet radius\\
tce\_max\_mult\_ev & TCE MES\\
ruwe & RUWE value\\
positional prob & positional probability\\
label & 0 for CFP, 1 for CP \\
ExoMiner score & \ExoMiner\ V1.2 score + priors \\
Boosted ExoMiner Score & score assigned by multiplicity \\ 
\bottomrule
\end{tabular}
\end{tabularx}
\end{threeparttable}
\end{table}
\endgroup

In order to provide insights into how the multiplicity boost improves the performance on specific scenarios, we provide in Table~\ref{table:performance_scenarios}, the total number of predicted FP for each scenario for \ExoMiner\ V1.2 and the multiplicity boost framework. As can be seen, multiplicity information helps improve the performance for individual scenarios.

\begin{table}[htb!]
 \centering
\caption{The performance of multiplicity boost framework compared to \ExoMiner\ V1.2 on individual scenarios in terms of FP counts.}
\label{table:scenarios}
\resizebox{\columnwidth}{!}{
\begin{threeparttable}
\begin{tabularx}{\linewidth}{@{}Y@{}}
\begin{tabular}{cccc}
\toprule
Scenarios & Observed & \ExoMiner\ V1.2 & Multiplicty Boost \\
\midrule
$[3, 0, 0]$ & 4 & 4 & 4 \\
$[1, 0, 0]$ & 90 & 89 & 89 \\
$[1, 0, 1]$ & 2 & 2 & 2 \\
$[1, 0, 2]$ & 0 & 0 & 0 \\
$[1, 1, 0]$ & 0 & 0 & 0 \\
$[1, 2, 2]$ & 0 & 0 & 0 \\
$[1, 4, 0]$ & 0 & 0 & 0 \\
\midrule
$[0, 0, 0]$ & 3406 & 3387 & 3395 \\
\midrule
$[0, 0, 1]$ & 23 & 25 & 23 \\
$[0, 0, 2]$ & 0 & 1 & 1 \\
$[0, 0, 3]$ & 0 & 0 & 0 \\
$[0, 1, 0]$ & 6 & 14 & 9 \\
$[0, 1, 1]$ & 0 & 2 & 2 \\
$[0, 1, 2]$ & 1 & 2 & 1 \\
$[0, 1, 3]$ & 0 & 0 & 0 \\
$[0, 2, 0]$ & 4 & 7 & 5 \\
$[0, 2, 1]$ & 0 & 0 & 0 \\
$[0, 2, 2]$ & 0 & 0 & 0 \\
$[0, 3, 0]$ & 0 & 1 & 0 \\
$[0, 3, 1]$ & 0 & 0 & 0 \\
$[0, 3, 2]$ & 1 & 1 & 0 \\
$[0, 4, 0]$ & 0 & 4 & 0 \\
$[0, 5, 0]$ & 1 & 0 & 0 \\
$[0, 6, 0]$ & 0 & 1 & 0 \\
\midrule
Total & 3538 & 3540 & 3531 \\
\bottomrule
\end{tabular}
\end{tabularx}
\end{threeparttable}
}
\label{table:performance_scenarios}
\end{table}

We also report in Table~\ref{table:classification_results} the scores of \ExoMiner\ V1.2 and the multiplicity logistic regression model on all 6181 CP and CFP KOIs discussed in Section~\ref{sec:dataset}. The multiplicity boost approach gives a score $>0.99$ to 1281 KOIs including one FP: K00082.06. This FP KOI is the 6th KOI in a system with five existing exoplanets. As mentioned in~\cite{Valizadegan_2022_ExoMiner}, this is incorrectly certified as FP in the Certified False Positive table. Given that there is a total of 1328 planets in multi-planet systems in \kepler\ Q1-Q17 DR25 data, the precision and recall value at the validation threshold of 0.99 are 1.000 and 0.965, respectively. The equivalent precision and recall values for the \ExoMiner\ V1.2 posterior without multiplicity boost are 1.0 and 0.836. Thus, the multiplicity boost framework significantly improves the recall at the validation threshold of 0.99.

\begin{figure}[ht]
\begin{center}
\centerline{\includegraphics[width=1.15\columnwidth]{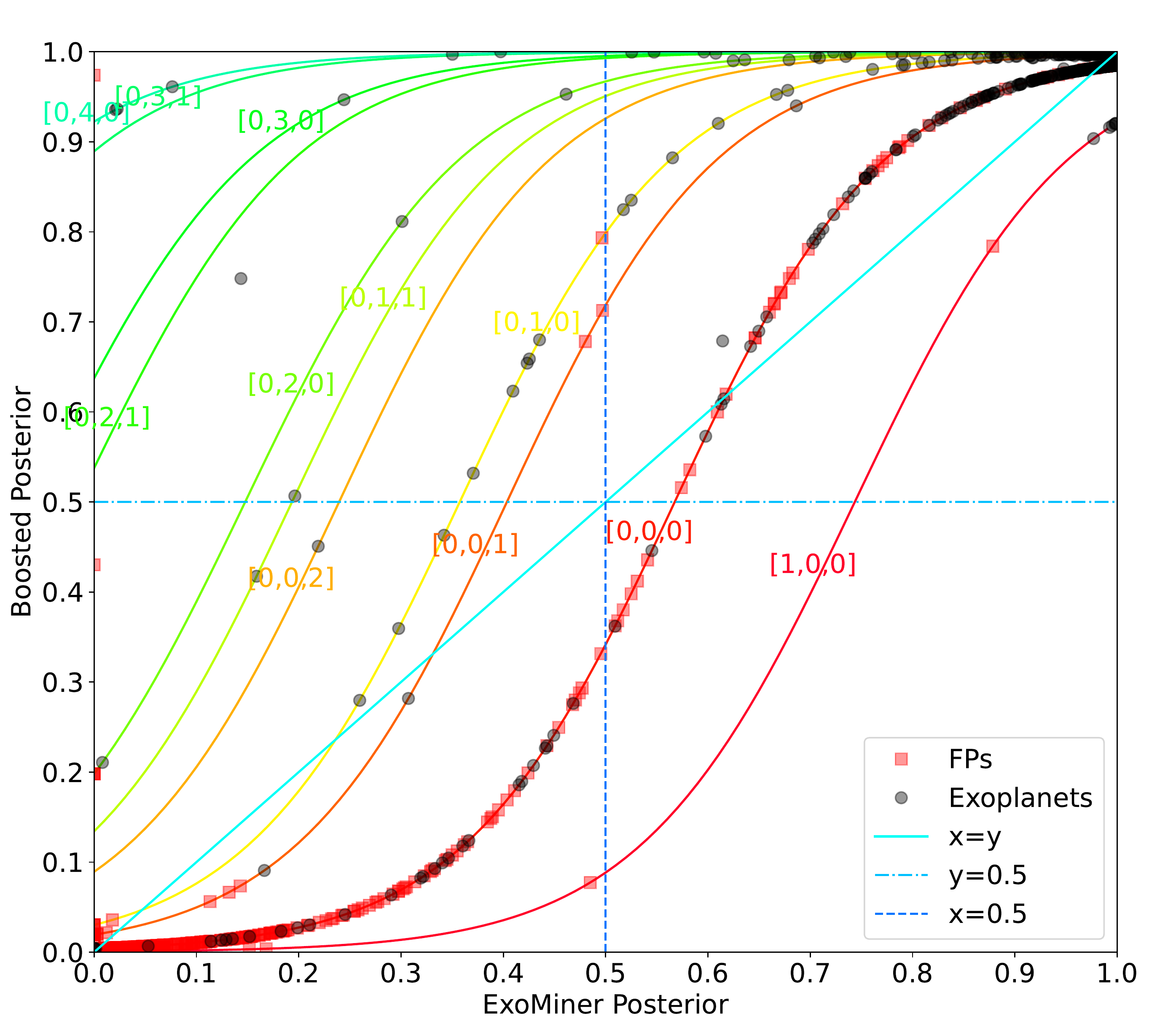}}
\caption{The score mapping of the existing exoplanets and FPs in the training data. We only show lines for tuples $\big[N_{FPs}(x),\; N_{CPs}(x),\; N_{UNKs}(x)\big]$ with more than ten points in the known KOI data for clarity. The lines for tuples with less than ten points exibit similar patterns.} 
\label{fig:training_points}
\end{center}
\end{figure}

To show how the known KOIs are mapped by our multiplicity boost framework, we plot the score mapping of the CPs and CFPs in the training set for the logistic regression model in Figure~\ref{fig:training_points}. We only show lines for tuples $\big[N_{FPs}(x),\; N_{CPs}(x),\; N_{UNKs}(x)\big]$ with more than ten points in the known KOI data. The mapping learned by the multiplicity boost classifier follows the general pattern of data seen in Table~\ref{table:scenarios}; i.e., the probability of CPs increases when the scenario has more existing CPs and unknown KOIs. Our multiplicity boost classifier also captures this pattern by increasing the scores for KOIs for target stars with already CPs and unknown KOIs. For single KOIs ($[0,0,0]$ curve), whose scenario includes the most points in this figure, even though the boosted score can change relative to the score of the original classifier (x-axis), it is always less than 0.99 and does not allow validation of any still unconfirmed single KOIs. 

\begingroup
\setlength{\tabcolsep}{2.5pt} 
\begin{table}[htb]
 \centering
\caption{The original scores of \ExoMiner\ V1.2 and the boosted scores after the application of multiplicity boost framework on 1570 unknown KOIs. This table describes the available columns. The full table is available online.}
\label{table:classification_results_unknowns}
\begin{threeparttable}
\begin{tabularx}{\linewidth}{@{}Y@{}}
\begin{tabular}{ll}
\toprule
Column & Description \\
\midrule
KIC & KIC ID\\
TCE & TCE planet number\\
KOI name & KOI name\\
Period (days) & TCE period\\
Radius (Re) & planet radius\\
tce\_max\_mult\_ev & TCE MES\\
ruwe & RUWE value\\
positional prob & positional probability\\
ExoMiner score & \ExoMiner\ V1.2 score + priors \\
Boosted ExoMiner Score & score assigned by multiplicity \\ 
\bottomrule
\end{tabular}
\end{tabularx}
\end{threeparttable}
\end{table}
\endgroup

\begin{table*}[htb!]
 \centering
\caption{Multiplicity scenarios and their total number of cases for 1570 unknown \kepler\ Q1-Q17 DR25 KOIs are reported in Column 1 and 2, respectively. The total number of KOIs with score $>0.99$, the number of KOIs that could be validated if their multiple event statistics (MES) $>10.5$, and validated exoplanets per each scenario are reported in Column 3, 4, and 5, respectively.}
\label{table:scenarios_unlabeled}
\begin{threeparttable}
\begin{tabularx}{\linewidth}{@{}Y@{}}
\begin{tabular}{ccccc}
\toprule
Scenario & Counts & Score $>0.99$ & Not validated due to MES $<10.5$ only & Validated Exoplanets \\
\midrule
$[0, 0, 0]$ & 1142  & 0   & 0 &0 \\
$[0, 0, 1]$ & 153  & 66   & 19 &27 \\
$[0, 0, 2]$ & 12  & 4   & 3 &1 \\
$[0, 0, 3]$ & 12  & 11   & 2 &5 \\
$[1, 0, 0]$ & 18  & 0   & 0 &0 \\
$[2, 0, 0]$ & 1  & 0   & 0 &0 \\
$[0, 1, 0]$ & 118  & 59   & 31 &15 \\
$[0, 1, 1]$ & 45  & 24   & 6 &8 \\
$[0, 1, 2]$ & 3  & 3   & 1 &2 \\
$[1, 1, 1]$ & 2  & 1   & 0 &1 \\
$[0, 2, 0]$ & 34  & 16   & 8 &7 \\
$[0, 2, 1]$ & 6  & 5   & 2 &1 \\
$[0, 2, 2]$ & 3  & 3   & 0 &0 \\
$[0, 3, 0]$ & 10  & 5   & 2 &2 \\
$[0, 3, 1]$ & 4  & 2   & 0 &0 \\
$[1, 3, 1]$ & 2  & 2   & 2 &0 \\
$[0, 4, 0]$ & 5  & 4   & 1 &0 \\
\midrule
Total & 1570 & 205 & 77 & 69 \\
\bottomrule
\end{tabular}
\end{tabularx}
\end{threeparttable}
\end{table*}

\subsection{Stability of New planets}
\label{sec:stability}
To ensure that the newly validated planets are stable, similar to~\cite{Dietrich_2020_hiddenworlds}, we use the dynamical stability parameter $\Delta$~\citep{Fabrycky_2014_multiplanets}, defined as:
\begin{equation}
    \Delta=\frac{2(a_2-a_1)}{a_2+a_1}\Big(\frac{3M^*}{m_1+m_2}\Big)^{1/3},
\end{equation}
where $a_1$ is the inner planet semi-major axis, $a_2$ is the outer semi-major axis, $M^*$ is the stellar mass, $m_1$ and $m_2$ are the planet masses, respectively. Similar to~\cite{Dietrich_2020_hiddenworlds}, we use $\Delta>8$ for stable systems.  

In order to use the above stability criterion, we need the planet mass, $M_p$, which is generally not available. To estimate the planet mass, we use the approach in~\cite{Fabrycky_2014_multiplanets} as:
\begin{equation}
    M_p=M_\Earth\big(\frac{R_p}{R_\Earth}\big)^\alpha
\end{equation}
where $\alpha=2.06$ for planet radius $R_p>R_\Earth$ and $\alpha=3$ for $R_p\leq R_\Earth$.

As a conservative approach, we also assume all unknown KOIs are confirmed planets when we calculate stability; i.e., the new planets should be stable when unknown KOIs are assumed to be planets.

\subsection{Vetoing criteria}
\label{sec:vetoes}
Given that the constructed training set for this study does not include FAs, we remove KOIs that have MES$<10.5$ as a conservative precaution to avoid noise-induced FAs. This threshold is similar to the one used by existing validation work~\cite{Rowe_2014_multiplanet, Morton-2016-vespa, armstrong-2020-exoplanet, Valizadegan_2022_ExoMiner}. It is actually based on the reliability analysis performed in~\citep{Thompson_2018}.~\cite{Thompson_2018} showed that the reliability meets the 0.99 threshold for MES $>10$ and period $<200$ days. For period $>200$ days reliability meets the 0.99 threshold when MES $>20$ and is 0.97 when $10<$ MES $<20$.

Also, because the multiplicity boost only works if the signal is known to be very close to the target, we veto KOIs from validation if 1) RUWE $>1.2$~\citep{Lindegren-2021-GAIA} or 2) positional probability $<0.99$~\citep{brysonKSCI2017} to prevent validation of KOIs for blended sources. This is because the multiplicity boost increases the score of the classifier significantly for scenarios with multiple planets and unknown KOIs.

In addition to the above vetoes, we make sure that \ExoMiner\ V1.2 score $>0.5$ to prevent validating planets that \ExoMiner\ V1.2 labels as FP. As another level of reliability, we remove any KOIs that have any FP flag set in the Cumulative KOI catalog.

\subsection{Newly Validated Exoplanets}
\label{sec:new_planets}
We applied \ExoMiner\ V1.2 and the multiplicity classifier trained in Section~\ref{sec:performance} to the 1570 unknown KOIs in the \kepler\ DR25 data set. Table~\ref{table:classification_results_unknowns} reports the \ExoMiner\ V1.2 and the multiplicity boost scores of all these 1570 unknown KOIs. This resulted in a total of 208 KOIs with boosted scores $>0.99$. Three out of these 208 KOIs were already confirmed by previous works but were not included in our 6181 known KOIs. They are K01831.03~\citep[Kepler-324 d, ][]{Jontof_Hutter_2021}, K00089.02~\citep[Kepler-462 c, ][]{Masuda_2020}, and K01783.02~\cite[Kepler-1662 c, ][]{Vissapragada_2020}. Out of these three, K01831.03 has a positional probability $<0.99$. 

Out of the remaining 205 KOIs, only K02926.05 has an original \ExoMiner\ score $<0.5$. K02926 has five previously confirmed planets and K02926.05 is the last KOI for this system in DR25. Besides the  original score veto, K02926.05 also has a positional probability $<0.99$. Regarding the KOIs' FP flags, there is only one KOI out of these 205 KOIs, K00408.05, that has the ``Not Transit-Like Flag'' set. The original \ExoMiner\ score of this KOI was 0.575 that was boosted to 0.999 given that there are four other CPs for this system. K00408.05 also has a low MES of 7.9. 

After removing KOIs with MES $<10.5$, RUWE $>1.2$, or positional probability $<0.99$, we validate a total of 69 new exoplanets, with their scenario counts presented in Table~\ref{table:scenarios_unlabeled}. All 69 of these planets satisfy the stability criteria discussed in Section~\ref{sec:stability}. Among these validated planets, we have two that have periods $>200$ days: K06103.01, whose MES is 35.4, and K01608.03, whose MES is 12.5. The former meets the reliability threshold of 0.99 (Check Section~\ref{sec:vetoes}). The latter has a reliability of 0.97 based on~\cite{Thompson_2018}. However, given that the reliability in~\cite{Thompson_2018} is calculated for single planet systems and given that K01608 has already two confirmed planets, the reliability of 0.97 is reasonable.

There are 77 KOIs that did not get validated simply because their MES $<10.5$. Note that out of 1570 unknown KOIs, 943 have MES $<10.5$. Thus, the majority of remaining unknown KOIs are in the low-MES region. This is due to the fact that almost all previous validation techniques avoid validating planets in this region. 

In general, the percentage of newly validated exoplanets increases for scenarios as the number of CPs or unknown KOIs increase. The exceptions to this rule are mainly due to our vetoing conditions preventing the validation of new exoplanets. For example, there are five unknown KOIs around target stars that already have four existing exoplanets. None of these  five unknown KOIs are validated; however, four of these five unknown KOIs have probability scores higher than 0.99 but fail at least one of the veto conditions. 

\begin{figure*}[ht]
\begin{center}
\centerline{\includegraphics[width=1.1\textwidth]{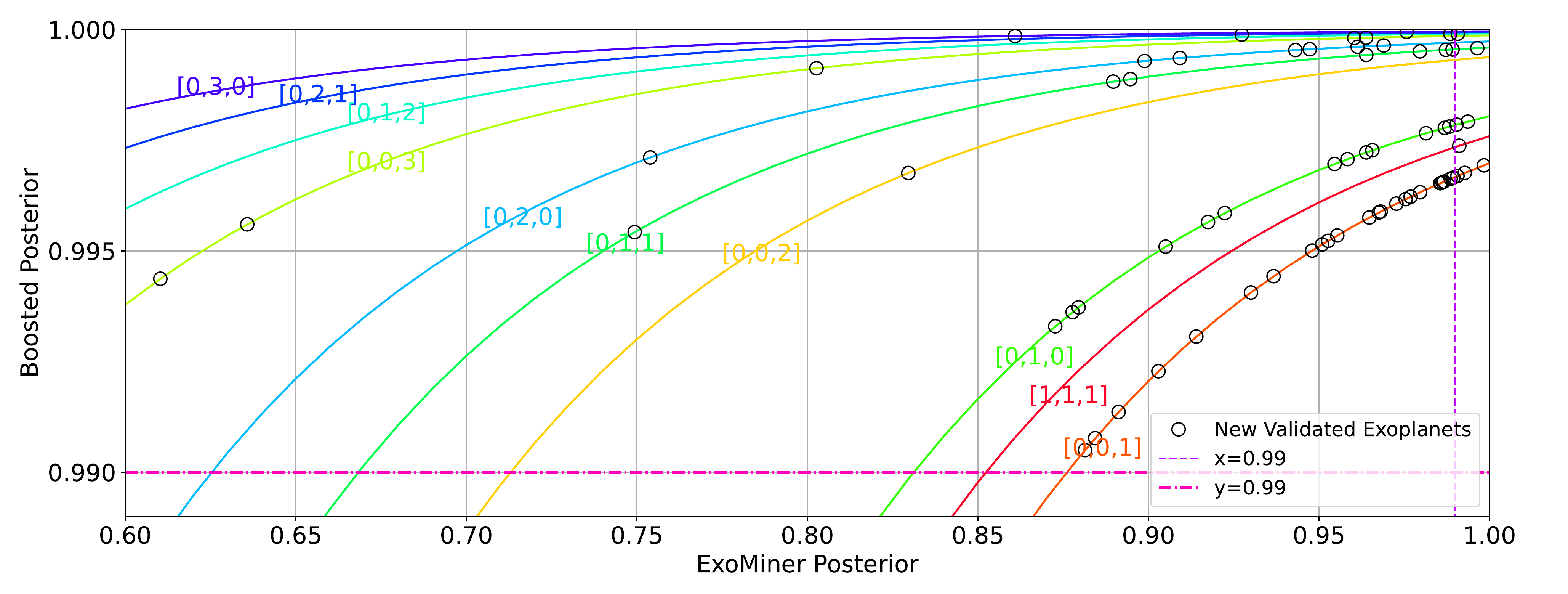}}
\caption{The newly validated exoplanets are shown using black circles. The mapping lines between \ExoMiner\ scores and the boosted scores for each tuple $\big[N_{FPs}(x),\; N_{CPs}(x),\; N_{UNKs}(x)\big]$ setting are displayed using solid lines. Lines are only plotted for tuples that are represented in the unknown KOIs.} 
\label{fig:validated_exoplanets}
\end{center}
\end{figure*}

Table~\ref{table:confirmedplanetlist} provides the list and properties of the 69 newly validated exoplanets. It also reports the probability score of \ExoMiner\ V1.2 after applying the prior and the probability score after using the logistic regression multiplicity boost approach. Note that the original score of \ExoMiner\ V1.2 for eight of these validated exoplanets is already larger than 0.99 so they could be validated even without multiplicity information. Figure~\ref{fig:validated_exoplanets} shows the score mapping done by our multiplicity boost model for these new exoplanets. We also report the list of KOIs with boosted score $>0.99$ that only failed the MES condition in Table~\ref{table:vetoed_list} and those that did not pass other vetoes in Table~\ref{table:vetoed_list2}.  

All of the newly validated planets are planet candidates in the Q1-Q17 DR25 KOI catalog according to the Exoplanet Archive disposition. We also manually examined the newly validated exoplanets to make sure there is nothing in the data indicating they should not be validated. Among the KOIs that do not have any existing CP, we discuss the following newly validated exoplanets, which are especially interesting. In several cases, the KOI failed the standard difference image centroiding test relative to the centroid of the OOT image, but passed the difference image centroiding test relative to the KIC catalog position of the target star. In these cases, the OOT centroid erroneously converged to a nearby star in the postage stamp rather than to the target star itself. The OOT centroid offset diagnostic result is incorrect when the OOT centroid does not locate the position of the target star \citep{Bryson_2013, Twicken_2018_DV}. Also note that none of these are flagged for ``Centroid Offset Flag'' or for any other FP flags in the Cumulative KOI catalog.
\begin{itemize}
    \item{K01358.01, K01358.02, K01358.03, and K01358.04:} We are validating all four KOIs of K01358 (represented by black circles on line [0,0,3] in Figure~\ref{fig:validated_exoplanets}). There is no FP or other KOI for this system. The minimum score of 0.635 for the KOIs in this system is assigned to K01358.01 by \ExoMiner\ V1.2. This score is boosted to 0.993 by the multiplicity boost information. All four candidates failed the difference image centroiding test relative to the out-of-transit centroid because the out-of-transit image centroid locked onto a much brighter (Kp = 13.5) star 6'' to the north of the Kp = 15.5 target star. The four candidates also failed the centroid shift test due to the presence of the brighter nearby star. However, all four passed the KIC offset difference image centroiding test, i.e., all four difference image centroids were consistent with the catalog position of the target star. 
    \item{K03145.01:} There are a total of three TCEs in DR25 for this system of which two are KOIs. Neither of these two KOIs are CPs or FPs and both appear to have difference image centroid positions relative to the catalog position of the host star consistent with being on target. The \ExoMiner\ V1.2 score for this KOI was 0.986 and boosted to 0.996 by the multiplicity boost information. The third (non-KOI) TCE corresponds to a 2\% deep (diluted) single transit or eclipse on a neighboring star $\sim$6 arcsec away.
    \item{K00102.01 and K00102.02:} There are two TCEs/KOIs in this system. We are validating both of them. This is a double star system with a separation of $\sim2.8$'' and a delta-magnitude of 1.5. In addition, there is a brighter star (KIC 8456687, Kp=10) 16'' south of the target star. Both KOIs fail the difference image centroid test relative to the OOT images due to the presence of this bright star. However, the difference image centroids relative to the KIC position for K00102.01 place the signal on the target star, ruling out the fainter companion at more than $30 \sigma$. The difference images for K00102.02 are much noisier due to the lower strength signal, however the difference images themselves indicate the transits fall on target.  Kepler photometry of KIC~8456687 shows no transit or eclipse-like signal at the 4~day period of K00102.02, ruling out the flux contamination from that target as the source. The Kepler follow-up observing program\footnote{\url{https://exofop.ipac.caltech.edu/}} carried out extensive observations on this early KOI and found no evidence of any additional companion brighter than a delta-magnitude of +7 at 3'' separation or greater. This indicates that a background source for this signal would require a stellar eclipse. However, the \ExoMiner\ V1.2 score for K00102.02 of 0.998 strongly favors a planetary transit. The \ExoMiner\ V1.2 scores for K00102.01 and K00102.02 were 0.986 and 0.998, which were boosted to 0.996 and 0.997, respectively.
    \item{K01276.01 and K01276.02:} There are two TCEs/KOIs for this system. We are validating both. There is a brighter star (KIC 8804292, Kp=13.3) $\sim$10 arc sec to the southeast of the target star that impacts the OOT centroids; however, the difference image centroids are consistent with being on target relative to the catalog star position. The \ExoMiner\ V1.2 scores for these KOIs were 0.977 and 0.989, which were boosted to 0.996 and 0.997, respectively.
    \item{K02612.01:} There are two TCEs/KOIs for this system. We are validating one. The target is a bright star, marginally saturated at Kp=11.8. The observed centroid offsets are along the brightest pixel column, consistent with the saturation bleed, and speckle and adaptive optics imaging from Kepler follow-up observations find no nearby companions. The \ExoMiner\ V1.2 score for this KOI was 0.968, which was boosted to 0.996.
    \item{K01871.01 and K01871.02:} There are two TCEs/KOIs for this system. We are validating both. This is a double star consisting of the target star (Kp=14.9) and a companion (KIC 9758087, Kp=14.6) $\sim5$'' to the northwest. The difference image centroids for both K01871.01 and K01871.02 are on the KIC position of the target star and rule out the companion at more than $10\sigma$. The \ExoMiner\ V1.2 scores for K01871.01 and K01871.02 were 0.955 and 0.947 respectively, which were both boosted to 0.995. 
    \item{K00253.01 and K00253.02:} This is a double star system of two Kp=15 stars with a separation of 5''. The second star (KIC 11752908) has a FP K02651.01 that triggered off of the signal from K00253.01. The difference image centroids for both K00253.01 and K00253.02 are within $\sim 0.5$'' of the KIC position of the target star and rule out the companion at more than $10\sigma$. The \ExoMiner\ V1.2 scores for these KOIs were 0.980 and 0.989, which were boosted to 0.996 and 0.997, respectively.
    \item{K02037.01:} There are three TCEs/KOIs for this system. While there is a nearby brighter star (KIC 9634819, Kp=14.2) 10'' to the north, the difference image centroids for all three KOIs are on the KIC position of the target star and rule out the companion. The \ExoMiner\ V1.2 score for this KOI was 0.830 which was boosted to 0.996 by multiplicity information given that there are two other unknown KOIs for this system (yellow line on Figure~\ref{fig:validated_exoplanets}).
\end{itemize}

As discussed in Section~\ref{sec:dataset}, K01082 is the only system with three unknown KOIs and one CP. We are validating K01082.01 and K01082.2 in this work. The only remaining KOI for K01082 not validated by this work is K01082.04, which has a score $>0.99$, but we decline to validate because of its low-MES value (MES=7.65). K01082.04 is listed in Table~\ref{table:vetoed_list}. Another interesting new validated planet is K03741.04. There are four TCEs/KOIs for K03741 which include one CP, one FP, and two unknown KOIs. This is the only planet we validate that has a host star with an FP (black circle on red line for scenario [1,1,1] in Figure~\ref{fig:validated_exoplanets}).

\begin{figure}[htb!]
	\centering
	\subfigure{\includegraphics[width=\columnwidth]{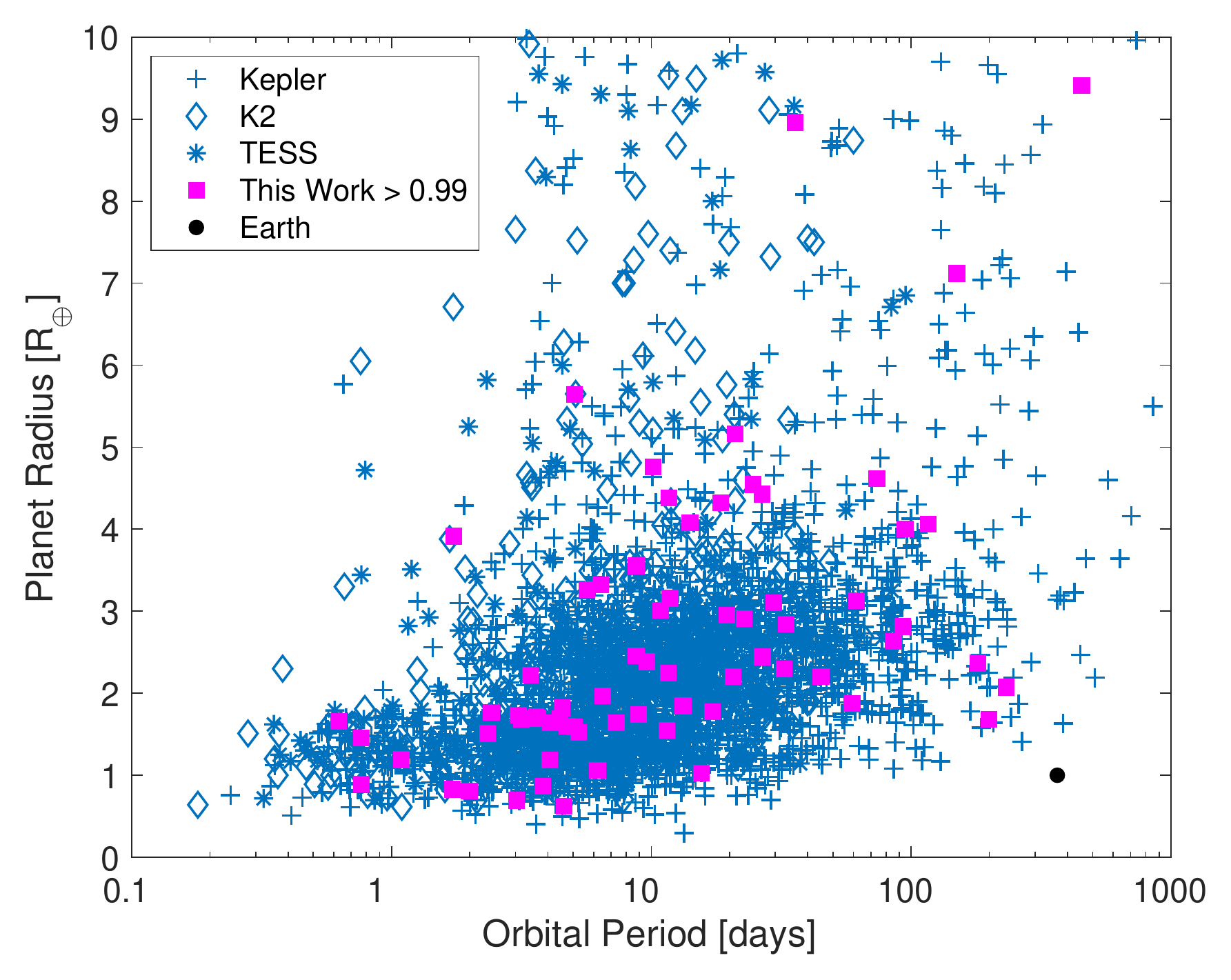}}	
\caption{Planet radius vs. orbital period for confirmed transiting planets and validated planets in this paper. The CPs are indicated by the discovery source with Kepler indicated by `+', K2 by a diamond `$\diamond$', and TESS by `*'. \ExoMiner\ V1.2-boosted validated planets are indicated by a magenta square. For reference, Earth is indicated by a black circle.}
\label{fig:radiusvsperiod}
\end{figure}

\begin{figure}[htb!]
	\centering
	\subfigure{\includegraphics[width=\columnwidth]{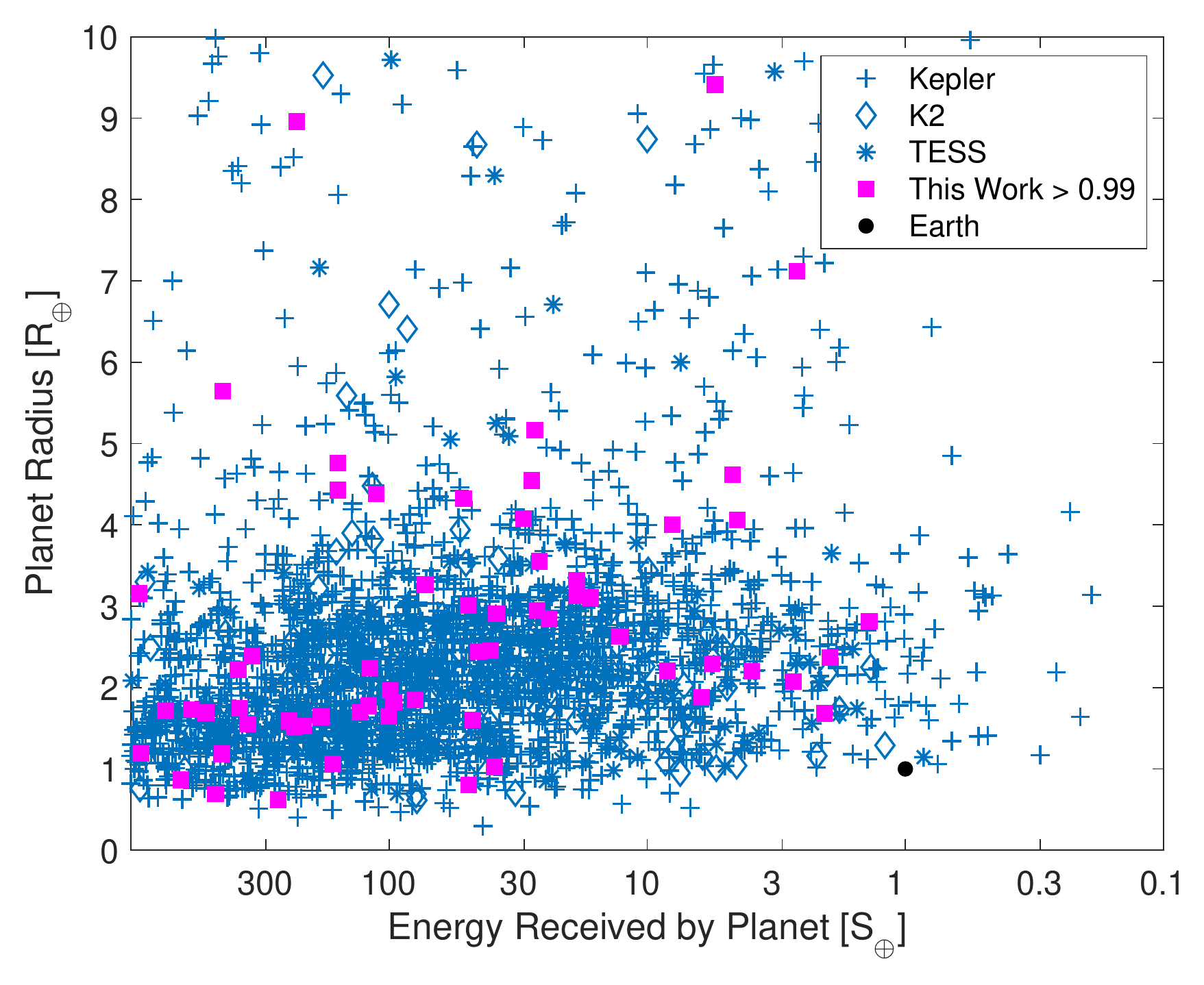}}	
\caption{Planet radius vs. energy received by planet for confirmed transiting planets and the validated planets in this paper. Symbols are the same as in Fig.~\ref{fig:radiusvsperiod}}
\label{fig:radiusvsenergy}
\end{figure}

Figure~\ref{fig:radiusvsperiod} displays a scatter plot of the planetary radius versus orbital period for the previously confirmed and validated exoplanets from the 
\kepler, K2 and TESS missions along with the 69 new \kepler\ exoplanets validated by our multiplicity framework. The distribution of the new exoplanets is consistent with that of the \kepler\ sample, with periods ranging from $\sim$0.6 days to over 450
days, and with planetary radii as small as $0.6\,R_\oplus$ to as large as $9.5\,R_\oplus$. Figure~\ref{fig:radiusvsenergy} shows a scatter plot of the planetary radius versus the energy received by each planet from the \kepler, K2, TESS missions and the new validated planets. As with the previous parametric plot, the distribution of the radius versus received energy of the new sample generally follows that of the \kepler\ sample.

\begin{table*}[htb!]
 \centering
\footnotesize
\caption{List of 69 newly validated exoplanets, sorted by TCE KIC.}
\label{table:confirmedplanetlist}
\resizebox{.85\textwidth}{!}{
\begin{threeparttable}
\begin{tabularx}{\linewidth}{@{}Y@{}}
\begin{tabular}{lcccccccccc}
\toprule
Number & TCE KIC & KOI Name & Period (days) & Radius (Re) & MES & Pos. Prob. & RUWE & \ExoMiner\ V1.2 & Multiplicity Score & \kepler\ Name \\
\midrule
\rownumber & 1717722.2 & K03145.01 &      4.54 &      1.82 &      13.8 &      1.000 &       0.96 &      0.986 &      0.997 & Kepler-1977 b\\
\rownumber & 2832589.1 & K01942.01 &     10.85 &      3.01 &      34.5 &      1.000 &       1.03 &      0.891 &      0.991 & Kepler-1978 b\\
\rownumber & 3229150.1 & K02150.01 &     18.51 &      4.32 &      22.5 &      1.000 &       1.08 &      0.965 &      0.996  & Kepler-1979 b\\
\rownumber & 3338885.2 & K01845.02 &      5.06 &      5.64 &      30.2 &      1.000 &       1.06 &      0.981 &      0.998  & Kepler-975 c\\
\rownumber & 3559860.1 & K03440.01 &     33.03 &      2.85 &      11.1 &      1.000 &       1.05 &      0.937 &      0.994  & Kepler-1980 b\\
\rownumber & 3561464.1 & K03398.02 &     35.80 &      8.96 &      15.3 &      1.000 &       0.99 &      0.895 &      0.999  & Kepler-1487 c\\
\rownumber & 3634051.1 & K06103.01 &    453.54 &      9.41 &      35.4 &      1.000 &       0.98 &      0.881 &      0.991  & Kepler-1981 b\\
\rownumber & 4077526.4 & K01336.04 &      4.46 &      1.69 &      11.4 &      1.000 &       1.00 &      0.861 &      1.000  & Kepler-58 e\\
\rownumber & 4157325.3 & K01860.03 &      3.08 &      1.73 &      22.4 &      1.000 &       0.95 &      0.927 &      1.000  & Kepler-416 d\\
\rownumber & 4173026.2 & K02172.02 &    116.58 &      4.06 &      15.0 &      1.000 &       0.99 &      0.990 &      0.998  & Kepler-1801 c\\
\rownumber & 4458082.2 & K02303.02 &      8.93 &      1.74 &      12.1 &      0.995 &       1.06 &      0.955 &      0.997  & Kepler-1181 c\\
\rownumber & 4548098.1 & K04157.01 &      3.82 &      0.87 &      13.4 &      1.000 &       0.95 &      0.991 &      0.997  & Kepler-1982 b\\
\rownumber & 4665571.1 & K02393.02 &      0.77 &      1.46 &      21.0 &      1.000 &       1.10 &      0.994 &      0.998  & Kepler-1834 c\\
\rownumber & 4770365.3 & K01475.03 &      4.73 &      1.60 &      10.9 &      1.000 &       1.03 &      0.899 &      0.999  & Kepler-1669 d\\
\rownumber & 4857058.1 & K03061.01 &      7.33 &      1.64 &      11.4 &      1.000 &       0.98 &      0.993 &      0.997  & Kepler-1983 b\\
\rownumber & 5531953.2 & K01681.02 &      1.99 &      0.81 &      12.9 &      1.000 &      -1.00 &      0.610 &      0.994  & Kepler-1984 b\\
\rownumber & 5942808.2 & K02250.02 &      0.63 &      1.65 &      23.5 &      1.000 &       1.02 &      0.987 &      0.998  & Kepler-1814 c\\
\rownumber & 6697605.1 & K02851.01 &      3.42 &      2.22 &      16.8 &      1.000 &      -1.00 &      0.986 &      0.997  & Kepler-1985 b\\
\rownumber & 7102227.3 & K01360.03 &      0.76 &      0.89 &      11.5 &      1.000 &       1.02 &      0.961 &      1.000  & Kepler-290 d\\
\rownumber & 7202957.1 & K02687.01 &      1.72 &      0.83 &      50.0 &      1.000 &       0.89 &      0.905 &      0.995  & Kepler-1869 c\\
\rownumber & 7285757.1 & K03271.01 &     19.55 &      2.95 &      11.1 &      1.000 &       1.01 &      0.951 &      0.995  & Kepler-1986 b\\
\rownumber & 7376983.1 & K01358.01 &      5.64 &      3.27 &      51.4 &      1.000 &       1.01 &      0.636 &      0.996  & Kepler-1987 d\\
\rownumber & 7376983.2 & K01358.02 &      8.74 &      2.45 &      25.0 &      1.000 &       1.01 &      0.803 &      0.999  & Kepler-1987 e\\
\rownumber & 7376983.3 & K01358.03 &      3.65 &      1.70 &      15.5 &      1.000 &       1.01 &      0.964 &      1.000  & Kepler-1987 c\\
\rownumber & 7376983.4 & K01358.04 &      2.35 &      1.51 &      14.6 &      1.000 &       1.01 &      0.960 &      1.000  & Kepler-1987 b\\
\rownumber & 7841925.2 & K01499.03 &      6.21 &      1.06 &      11.0 &      1.000 &       0.91 &      0.890 &      0.999  & Kepler-865 c\\
\rownumber & 7869917.2 & K01525.02 &     11.81 &      3.16 &      12.1 &      0.998 &       0.94 &      0.964 &      0.997  & Kepler-880 c\\
\rownumber & 7939330.1 & K01581.01 &     29.54 &      3.11 &      31.4 &      1.000 &       1.00 &      0.988 &      0.998  & Kepler-896 c\\
\rownumber & 7983117.1 & K03214.01 &     11.49 &      1.55 &      15.8 &      1.000 &      -1.00 &      0.987 &      0.997  & Kepler-1988 b\\
\rownumber & 8162789.1 & K00521.01 &     10.16 &      4.76 &     102.0 &      1.000 &       0.98 &      0.903 &      0.992  & Kepler-1989 b\\
\rownumber & 8456679.1 & K00102.01 &      1.74 &      3.91 &     432.9 &      1.000 &       0.88 &      0.986 &      0.997  & Kepler-1990 b\\
\rownumber & 8456679.2 & K00102.02 &      4.07 &      1.19 &      22.5 &      1.000 &       0.88 &      0.998 &      0.997  & Kepler-1990 c\\
\rownumber & 8780959.3 & K03741.04 &      9.63 &      2.38 &      19.7 &      1.000 &       0.96 &      0.991 &      0.997  & Kepler-1518 c\\
\rownumber & 8804283.1 & K01276.01 &     22.79 &      2.91 &      40.1 &      1.000 &       0.93 &      0.977 &      0.996  & Kepler-1991 c\\
\rownumber & 8804283.2 & K01276.02 &     13.26 &      1.85 &      14.6 &      1.000 &       0.93 &      0.989 &      0.997  & Kepler-1991 b\\
\rownumber & 8827575.2 & K03052.02 &     15.61 &      1.03 &      11.3 &      1.000 &       1.07 &      0.953 &      0.995  & Kepler-1992 b\\
\rownumber & 8890924.1 & K04269.01 &     26.66 &      4.43 &      11.7 &      1.000 &       1.00 &      0.914 &      0.993  & Kepler-1993 b\\
\rownumber & 9071593.2 & K02257.02 &     59.28 &      1.88 &      10.7 &      1.000 &       1.09 &      0.878 &      0.994  & Kepler-1162 c\\
\rownumber & 9100953.2 & K04500.02 &     44.99 &      2.21 &      10.7 &      1.000 &       1.03 &      0.980 &      1.000  & Kepler-1610 c\\
\rownumber & 9117416.2 & K03425.02 &      3.16 &      1.69 &      13.8 &      1.000 &       1.14 &      0.987 &      1.000  & Kepler-1921 c\\
\rownumber & 9205938.2 & K02162.02 &    199.67 &      1.68 &      11.1 &      1.000 &       0.91 &      0.922 &      0.996  & Kepler-1126 c\\
\rownumber & 9285265.2 & K03410.02 &     61.57 &      3.13 &      10.9 &      1.000 &       1.11 &      0.958 &      0.997  & Kepler-1491 c\\
\rownumber & 9602613.1 & K02612.01 &      4.61 &      0.62 &      11.4 &      1.000 &       0.96 &      0.968 &      0.996  & Kepler-1994 b\\
\rownumber & 9634821.1 & K02037.01 &     73.76 &      4.62 &      29.9 &      1.000 &       1.12 &      0.830 &      0.997  & Kepler-1995 b\\
\rownumber & 9636135.2 & K01498.02 &      2.42 &      1.77 &      13.3 &      1.000 &       0.90 &      0.966 &      0.997  & Kepler-864 c\\
\rownumber & 9729691.2 & K01751.02 &     21.00 &      5.16 &      27.8 &      1.000 &       0.97 &      0.873 &      0.993  & Kepler-949 c\\
\rownumber & 9758089.1 & K01871.01 &     92.73 &      2.81 &      31.7 &      1.000 &       1.08 &      0.955 &      0.995  & Kepler-1996 c \\
\rownumber & 9758089.2 & K01871.02 &     32.38 &      2.29 &      26.3 &      1.000 &       1.08 &      0.948 &      0.995  & Kepler-1996 b\\
\rownumber & 9785921.1 & K03372.01 &     26.76 &      2.44 &      11.1 &      0.993 &       1.02 &      0.975 &      0.996  & Kepler-1997 b\\
\rownumber & 9839821.2 & K02012.02 &    180.93 &      2.37 &      12.0 &      0.999 &       0.97 &      0.917 &      0.996  & Kepler-1052 c\\
\rownumber & 9896018.2 & K02579.02 &      3.60 &      1.72 &      12.8 &      1.000 &       1.00 &      0.964 &      0.999  & Kepler-1859 c\\
\rownumber & 10055126.3 & K01608.03 &    232.05 &      2.07 &      12.5 &      1.000 &       1.12 &      0.909 &      0.999 & Kepler-311 d\\
\rownumber & 10141900.1 & K01082.01 &      6.50 &      1.97 &      15.9 &      1.000 &       0.91 &      0.991 &      1.000 & Kepler-763 d\\
\rownumber & 10141900.3 & K01082.02 &      4.10 &      1.64 &      11.0 &      1.000 &       0.91 &      0.988 &      1.000 & Kepler-763 c\\
\rownumber & 10189546.1 & K00427.01 &     24.61 &      4.55 &      53.9 &      1.000 &       1.15 &      0.969 &      1.000 & Kepler-549 d\\
\rownumber & 10265898.2 & K00732.03 &      5.25 &      1.52 &      10.6 &      1.000 &       1.01 &      0.749 &      0.995 & Kepler-656 c\\
\rownumber & 10350571.1 & K01175.02 &     17.16 &      1.77 &      13.0 &      1.000 &       1.01 &      0.996 &      1.000 & Kepler-784 c\\
\rownumber & 10387742.1 & K02583.01 &      3.03 &      0.69 &      12.9 &      1.000 &       1.01 &      0.968 &      0.996 & Kepler-1998 b\\
\rownumber & 10460984.3 & K00474.03 &     94.89 &      4.00 &      19.5 &      1.000 &       1.10 &      0.976 &      1.000 & Kepler-164 e\\
\rownumber & 10843431.1 & K07378.01 &      8.74 &      3.55 &      12.1 &      1.000 &       1.00 &      0.884 &      0.991 & Kepler-1999 b\\
\rownumber & 10973664.2 & K00601.02 &     11.68 &      4.38 &      43.0 &      1.000 &       1.03 &      0.943 &      1.000 & Kepler-618 d\\
\rownumber & 11122894.2 & K01426.03 &    150.02 &      7.12 &      92.4 &      1.000 &       0.94 &      0.754 &      0.997 & Kepler-297 d\\
\rownumber & 11450414.3 & K01992.03 &     85.52 &      2.63 &      11.2 &      1.000 &       1.02 &      0.947 &      1.000 & Kepler-347 d\\
\rownumber & 11618601.2 & K03022.02 &      5.05 &      1.59 &      10.6 &      0.999 &       1.05 &      0.879 &      0.994 & Kepler-1894 c\\
\rownumber & 11752906.1 & K00253.01 &      6.38 &      3.32 &      55.2 &      1.000 &       1.01 &      0.980 &      0.996 & Kepler-2000 b\\
\rownumber & 11752906.2 & K00253.02 &     20.62 &      2.20 &      12.3 &      1.000 &       1.01 &      0.989 &      0.997 & Kepler-2000 c\\
\rownumber & 11810124.2 & K03344.01 &     11.60 &      2.24 &      11.7 &      1.000 &       1.14 &      0.989 &      1.000 & Kepler-1471 c\\
\rownumber & 12061969.1 & K02061.01 &     14.09 &      4.08 &      15.9 &      1.000 &       0.94 &      0.930 &      0.994 & Kepler-2001 c\\
\rownumber & 12061969.2 & K02061.02 &      1.09 &      1.19 &      10.5 &      1.000 &       0.94 &      0.973 &      0.996 & Kepler-2001 b\\
\bottomrule
\end{tabular}
\end{tabularx}
\end{threeparttable}
}
\end{table*}

\begin{table*}[htb!]
 \centering
\footnotesize
\caption{List of low-MES KOIs (MES$<10.5$) with boosted \ExoMiner\ V1.2 score $>0.99$, RUWE $<1.2$, and positional probability $>0.99$. This list is sorted by MES value.}
\label{table:vetoed_list}
\resizebox{.77\textwidth}{!}{
\begin{threeparttable}
\begin{tabularx}{\linewidth}{@{}Y@{}}
\begin{tabular}{lcccccccccc}
\toprule
Number & TCE KIC & KOI Name & Period (days) & Radius (Re) & MES & Pos. Prob. & RUWE & \ExoMiner\ V1.2 & Multiplicity Score  \\
\midrule
\rownumbervetoed & 4860678.2 & K01602.02 &      3.03 &      1.83 &      10.4 &      1.000 &       0.94 &      0.960 &      0.997  \\
\rownumbervetoed & 6209677.2 & K01750.02 &      2.54 &      1.01 &      10.3 &      1.000 &       1.08 &      0.854 &      0.992  \\
\rownumbervetoed & 10717241.2 & K00430.02 &      9.34 &      0.98 &      10.2 &      1.000 &       1.05 &      0.831 &      0.990  \\
\rownumbervetoed & 5903749.1 & K03029.01 &     18.98 &      2.49 &      10.2 &      1.000 &      -1.00 &      0.996 &      0.997  \\
\rownumbervetoed & 4164922.2 & K03864.02 &     18.26 &      1.07 &      10.2 &      1.000 &       1.04 &      0.999 &      0.998  \\
\rownumbervetoed & 7449554.2 & K02357.02 &     15.90 &      4.82 &      10.2 &      1.000 &       1.11 &      0.994 &      0.997  \\
\rownumbervetoed & 3230805.1 & K03068.01 &      3.92 &      1.00 &      10.2 &      1.000 &       0.94 &      0.986 &      0.997  \\
\rownumbervetoed & 7692093.1 & K03337.01 &     11.17 &      1.66 &      10.2 &      1.000 &       1.02 &      0.997 &      0.997  \\
\rownumbervetoed & 8874090.2 & K01404.02 &     18.91 &      1.19 &      10.1 &      1.000 &       1.06 &      0.997 &      0.998  \\
\rownumbervetoed & 9002538.2 & K03196.02 &      6.88 &      0.81 &      10.1 &      1.000 &       0.95 &      0.996 &      0.997  \\
\rownumbervetoed & 8760040.2 & K02963.02 &      7.35 &      1.54 &      10.1 &      1.000 &       1.07 &      0.991 &      0.998  \\
\rownumbervetoed & 9489524.4 & K02029.04 &      4.79 &      0.78 &       9.9 &      1.000 &       1.08 &      0.998 &      1.000  \\
\rownumbervetoed & 5511659.1 & K04541.01 &      8.53 &      2.84 &       9.9 &      1.000 &       0.94 &      0.987 &      0.997  \\
\rownumbervetoed & 8644365.2 & K03384.01 &     10.55 &      1.30 &       9.9 &      1.000 &       1.08 &      0.889 &      0.994  \\
\rownumbervetoed & 9836563.1 & K04421.01 &      4.73 &      0.73 &       9.8 &      1.000 &       0.96 &      0.883 &      0.991  \\
\rownumbervetoed & 5621333.2 & K03341.02 &     11.55 &      1.66 &       9.8 &      1.000 &       1.04 &      0.950 &      0.995  \\
\rownumbervetoed & 6436505.1 & K06707.01 &     43.54 &      1.82 &       9.8 &      1.000 &       1.02 &      0.993 &      0.997  \\
\rownumbervetoed & 6871071.4 & K02220.04 &      7.66 &      1.83 &       9.8 &      1.000 &       0.94 &      0.998 &      1.000  \\
\rownumbervetoed & 9093086.1 & K06191.01 &      9.70 &      1.67 &       9.7 &      1.000 &       0.99 &      0.953 &      0.995  \\
\rownumbervetoed & 4472818.2 & K03878.02 &     15.36 &      1.20 &       9.7 &      1.000 &       0.89 &      0.889 &      0.994  \\
\rownumbervetoed & 7289317.2 & K02450.02 &      7.19 &      1.32 &       9.7 &      1.000 &       0.96 &      0.996 &      0.998  \\
\rownumbervetoed & 4645174.2 & K03437.02 &     34.75 &      2.53 &       9.6 &      1.000 &       1.04 &      0.962 &      0.997  \\
\rownumbervetoed & 7825899.3 & K00896.03 &     28.87 &      1.70 &       9.6 &      1.000 &       0.92 &      0.994 &      1.000  \\
\rownumbervetoed & 6705026.2 & K03374.02 &     34.11 &      2.09 &       9.5 &      1.000 &       1.03 &      0.895 &      0.995  \\
\rownumbervetoed & 7289338.1 & K03420.02 &     12.50 &      2.73 &       9.5 &      1.000 &       0.94 &      0.990 &      0.997  \\
\rownumbervetoed & 4770617.3 & K02243.03 &     31.45 &      3.42 &       9.4 &      1.000 &       0.99 &      0.787 &      0.998  \\
\rownumbervetoed & 9967009.1 & K03462.01 &     12.43 &      2.01 &       9.4 &      1.000 &       0.98 &      0.938 &      0.994  \\
\rownumbervetoed & 8168187.2 & K02209.02 &     35.50 &      2.10 &       9.4 &      1.000 &       1.03 &      0.886 &      0.994  \\
\rownumbervetoed & 2854181.2 & K02232.02 &     12.84 &      1.95 &       9.3 &      1.000 &       0.97 &      0.989 &      0.998  \\
\rownumbervetoed & 9896018.3 & K02579.03 &     10.30 &      1.86 &       9.3 &      1.000 &       1.00 &      0.996 &      1.000  \\
\rownumbervetoed & 5003670.2 & K04524.02 &      4.53 &      1.23 &       9.3 &      1.000 &       0.86 &      0.887 &      0.999  \\
\rownumbervetoed & 11968463.4 & K02433.04 &     27.90 &      2.11 &       9.2 &      0.999 &       0.94 &      0.982 &      1.000  \\
\rownumbervetoed & 7106173.1 & K03083.01 &     10.18 &      0.92 &       9.2 &      1.000 &       0.98 &      0.871 &      0.998  \\
\rownumbervetoed & 12256520.2 & K02264.02 &      7.25 &      1.15 &       9.2 &      1.000 &       0.87 &      0.857 &      0.992  \\
\rownumbervetoed & 5769810.1 & K04913.02 &      8.97 &      3.39 &       9.2 &      0.999 &       1.12 &      0.900 &      0.992  \\
\rownumbervetoed & 6508221.3 & K00416.03 &      9.75 &      1.54 &       9.2 &      1.000 &       0.95 &      0.792 &      0.998  \\
\rownumbervetoed & 11918099.2 & K00780.02 &      7.24 &      1.81 &       9.2 &      1.000 &       0.96 &      0.979 &      0.998  \\
\rownumbervetoed & 4735826.2 & K03184.03 &      4.02 &      0.67 &       9.1 &      1.000 &       0.90 &      0.974 &      0.996  \\
\rownumbervetoed & 9655711.1 & K06209.01 &     12.71 &      2.29 &       9.1 &      1.000 &       0.97 &      0.942 &      0.995  \\
\rownumbervetoed & 7285757.2 & K03271.02 &      7.42 &      1.26 &       9.0 &      1.000 &       1.01 &      0.992 &      0.997  \\
\rownumbervetoed & 5177859.2 & K04246.02 &      8.76 &      1.36 &       9.0 &      1.000 &       1.10 &      0.985 &      0.998  \\
\rownumbervetoed & 1432789.2 & K00992.02 &      4.58 &      1.64 &       8.9 &      1.000 &       1.00 &      0.903 &      0.995  \\
\rownumbervetoed & 10290666.2 & K00332.02 &      6.87 &      0.88 &       8.8 &      1.000 &       1.14 &      0.977 &      0.998  \\
\rownumbervetoed & 4157325.4 & K01860.04 &     24.84 &      1.57 &       8.8 &      1.000 &       0.95 &      0.575 &      0.997  \\
\rownumbervetoed & 6543893.3 & K01627.03 &      3.81 &      3.28 &       8.8 &      1.000 &       1.03 &      0.811 &      0.998  \\
\rownumbervetoed & 7047922.2 & K01899.02 &     10.52 &      1.19 &       8.7 &      1.000 &       1.13 &      0.947 &      0.997  \\
\rownumbervetoed & 10000941.2 & K04146.02 &      2.57 &      0.69 &       8.7 &      1.000 &       1.07 &      0.942 &      0.997  \\
\rownumbervetoed & 1996180.2 & K02534.02 &      5.42 &      1.35 &       8.7 &      1.000 &       1.07 &      0.866 &      0.993  \\
\rownumbervetoed & 7107802.2 & K02420.02 &      5.47 &      1.19 &       8.6 &      1.000 &       1.05 &      0.979 &      0.998  \\
\rownumbervetoed & 5903749.2 & K03029.02 &      6.35 &      1.42 &       8.6 &      1.000 &      -1.00 &      0.916 &      0.993  \\
\rownumbervetoed & 9100953.3 & K04500.03 &     14.75 &      2.01 &       8.6 &      1.000 &       1.03 &      0.953 &      0.999  \\
\rownumbervetoed & 12266636.2 & K01522.02 &     12.65 &      1.14 &       8.5 &      1.000 &       0.98 &      0.914 &      0.996  \\
\rownumbervetoed & 7663405.2 & K01519.02 &     57.13 &      1.93 &       8.5 &      1.000 &       1.02 &      0.975 &      0.998  \\
\rownumbervetoed & 9872283.2 & K01815.02 &      1.75 &      1.56 &       8.4 &      1.000 &       1.10 &      0.844 &      0.991  \\
\rownumbervetoed & 7512982.2 & K01480.02 &      7.00 &      1.47 &       8.4 &      1.000 &       0.99 &      0.975 &      0.998  \\
\rownumbervetoed & 6276791.2 & K04477.01 &      5.30 &      1.28 &       8.4 &      1.000 &       1.04 &      0.872 &      0.993  \\
\rownumbervetoed & 8613535.4 & K02263.03 &     15.59 &      1.15 &       8.4 &      1.000 &       0.98 &      0.740 &      0.997  \\
\rownumbervetoed & 9602613.2 & K02612.02 &      7.57 &      0.61 &       8.3 &      1.000 &       0.96 &      0.895 &      0.992  \\
\rownumbervetoed & 7021681.2 & K00255.02 &     13.60 &      0.79 &       8.3 &      1.000 &       1.03 &      0.965 &      0.999  \\
\rownumbervetoed & 10265898.3 & K00732.02 &      3.30 &      1.44 &       8.2 &      1.000 &       1.01 &      0.966 &      0.999  \\
\rownumbervetoed & 3645438.3 & K04385.03 &     17.37 &      1.83 &       8.2 &      1.000 &       1.05 &      0.827 &      0.999  \\
\rownumbervetoed & 7289338.2 & K03420.01 &      5.77 &      2.10 &       8.2 &      1.000 &       0.94 &      0.950 &      0.995  \\
\rownumbervetoed & 6221385.3 & K06145.03 &      7.31 &      2.41 &       8.2 &      1.000 &       1.08 &      0.947 &      1.000  \\
\rownumbervetoed & 8216763.1 & K04838.01 &     13.30 &      0.95 &       8.1 &      1.000 &       1.00 &      0.808 &      0.996  \\
\rownumbervetoed & 7673841.3 & K02585.03 &      7.88 &      0.90 &       8.1 &      1.000 &       0.94 &      0.824 &      0.999  \\
\rownumbervetoed & 8216763.2 & K04838.03 &     24.07 &      0.97 &       8.1 &      1.000 &       1.00 &      0.961 &      0.999  \\
\rownumbervetoed & 6103377.2 & K03004.02 &      7.04 &      2.81 &       8.1 &      1.000 &       1.06 &      0.974 &      0.997  \\
\rownumbervetoed & 5531953.3 & K01681.04 &     21.91 &      1.05 &       8.0 &      1.000 &      -1.00 &      0.919 &      1.000  \\
\rownumbervetoed & 5351250.5 & K00408.05 &     93.80 &      2.48 &       7.9 &      1.000 &       1.01 &      0.576 &      1.000  \\
\rownumbervetoed & 8226050.2 & K01910.02 &     18.38 &      1.36 &       7.8 &      1.000 &       1.02 &      0.874 &      0.993  \\
\rownumbervetoed & 5602588.2 & K02369.03 &      7.23 &      1.55 &       7.7 &      1.000 &       0.96 &      0.948 &      0.997  \\
\rownumbervetoed & 6716545.2 & K02906.03 &     21.94 &      1.37 &       7.7 &      1.000 &       1.05 &      0.908 &      0.999  \\
\rownumbervetoed & 7765528.2 & K01840.02 &      9.39 &      1.94 &       7.7 &      1.000 &       0.96 &      0.875 &      0.993  \\
\rownumbervetoed & 10141900.4 & K01082.04 &      9.66 &      1.77 &       7.7 &      1.000 &       0.91 &      0.559 &      0.994  \\
\rownumbervetoed & 5531953.4 & K01681.03 &      3.53 &      0.88 &       7.6 &      1.000 &      -1.00 &      0.977 &      1.000  \\
\rownumbervetoed & 10028792.4 & K01574.04 &      8.98 &      1.84 &       7.4 &      1.000 &       0.90 &      0.896 &      1.000  \\
\rownumbervetoed & 11968463.6 & K02433.07 &     86.43 &      2.98 &       7.2 &      1.000 &       0.94 &      0.901 &      1.000  \\
\bottomrule
\end{tabular}
\end{tabularx}
\end{threeparttable}
}
\end{table*}

\begin{table*}
 \centering
\footnotesize
\caption{List of KOIs with boosted \ExoMiner\ V1.2 score $>0.99$ that did not pass positional probability or RUWE tests. This list is sorted by positional probability.}
\label{table:vetoed_list2}
\resizebox{.85\textwidth}{!}{
\begin{threeparttable}
\begin{tabularx}{\linewidth}{@{}Y@{}}
\begin{tabular}{lcccccccccc}
\toprule
Number & TCE KIC & KOI Name & Period (days) & Radius (Re) & MES & Pos. Prob. & RUWE & \ExoMiner\ V1.2 & Multiplicity Score  \\
\midrule
\rownumbervetoedfull & 10662202.3 & K00750.03 &     14.52 &      1.55 &       7.5 &      1.000 &       1.36 &      0.914 &      0.999  \\
\rownumbervetoedfull & 10397751.3 & K02859.05 &      5.43 &      0.76 &       9.4 &      1.000 &       1.65 &      0.949 &      1.000  \\
\rownumbervetoedfull & 10397751.4 & K02859.04 &      2.91 &      0.53 &       8.6 &      1.000 &       1.65 &      0.988 &      1.000  \\
\rownumbervetoedfull & 8280511.4 & K01151.04 &     17.45 &      0.87 &       8.4 &      1.000 &       2.01 &      0.948 &      1.000  \\
\rownumbervetoedfull & 8280511.5 & K01151.05 &     21.72 &      0.92 &       7.8 &      1.000 &       2.01 &      0.871 &      1.000  \\
\rownumbervetoedfull & 10875007.2 & K04149.02 &     14.71 &      1.68 &      10.6 &      1.000 &       2.05 &      0.916 &      0.993  \\
\rownumbervetoedfull & 8261920.1 & K02174.01 &      6.69 &      2.66 &      17.9 &      1.000 &       2.82 &      0.975 &      1.000  \\
\rownumbervetoedfull & 5956656.2 & K01053.02 &     46.25 &      2.48 &       8.8 &      1.000 &       3.33 &      0.984 &      0.998  \\
\rownumbervetoedfull & 5856571.2 & K01839.02 &     80.41 &      4.41 &      12.9 &      1.000 &       3.85 &      0.929 &      0.996  \\
\rownumbervetoedfull & 3529290.1 & K03340.02 &     13.73 &      1.08 &      11.4 &      1.000 &       4.09 &      0.919 &      0.993  \\
\rownumbervetoedfull & 5629353.1 & K06132.01 &     33.32 &     13.31 &      59.2 &      1.000 &       4.39 &      0.787 &      0.997  \\
\rownumbervetoedfull & 5542466.3 & K01590.03 &      4.75 &      1.63 &       9.6 &      1.000 &       5.07 &      0.988 &      1.000  \\
\rownumbervetoedfull & 4832837.2 & K00605.02 &      5.07 &      0.73 &       9.5 &      1.000 &      10.81 &      0.973 &      0.997  \\
\rownumbervetoedfull & 4736569.2 & K01996.02 &      7.07 &      1.09 &       9.1 &      1.000 &      10.88 &      0.969 &      0.997  \\
\rownumbervetoedfull & 11566064.1 & K00353.01 &    152.11 &     11.67 &      82.5 &      1.000 &       1.42 &      0.833 &      0.998  \\
\rownumbervetoedfull & 5384713.1 & K03444.02 &     60.33 &      5.25 &      59.2 &      1.000 &       1.95 &      0.959 &      1.000  \\
\rownumbervetoedfull & 11125797.1 & K03371.01 &     58.13 &      1.55 &      10.8 &      1.000 &       2.90 &      0.964 &      0.997  \\
\rownumbervetoedfull & 9413156.1 & K04700.01 &      3.83 &      1.32 &       9.3 &      0.981 &       0.90 &      0.903 &      0.992  \\
\rownumbervetoedfull & 4385148.1 & K02942.01 &     13.84 &      2.40 &      14.5 &      0.975 &       0.92 &      0.982 &      0.998  \\
\rownumbervetoedfull & 5031857.2 & K01573.02 &      7.14 &      1.31 &      13.6 &      0.974 &       1.08 &      0.954 &      0.997  \\
\rownumbervetoedfull & 5080636.2 & K01843.02 &      6.36 &      0.69 &       9.9 &      0.962 &       1.19 &      0.851 &      0.992  \\
\rownumbervetoedfull & 7100673.5 & K04032.05 &      7.24 &      0.88 &       9.0 &      0.961 &       1.19 &      0.986 &      1.000  \\
\rownumbervetoedfull & 8581240.1 & K03111.01 &     10.77 &      1.05 &      11.9 &      0.959 &       1.07 &      0.990 &      0.997  \\
\rownumbervetoedfull & 4466677.2 & K01338.02 &     42.04 &      1.73 &      12.1 &      0.950 &       0.96 &      0.934 &      0.999  \\
\rownumbervetoedfull & 4478168.2 & K00626.02 &      8.03 &      1.13 &       9.9 &      0.941 &       0.90 &      0.985 &      0.998  \\
\rownumbervetoedfull & 10028792.3 & K01574.03 &      5.83 &      1.79 &       9.3 &      0.940 &       0.90 &      0.750 &      0.999  \\
\rownumbervetoedfull & 9002538.1 & K03196.01 &      4.96 &      0.71 &      11.0 &      0.938 &       0.95 &      0.967 &      0.996  \\
\rownumbervetoedfull & 6026737.1 & K02949.01 &     10.17 &      1.15 &      10.0 &      0.925 &       0.89 &      0.997 &      0.997  \\
\rownumbervetoedfull & 10122538.5 & K02926.05 &     75.73 &      3.55 &      11.4 &      0.917 &       1.06 &      0.356 &      0.997  \\
\rownumbervetoedfull & 7449554.1 & K02357.01 &      2.42 &      2.64 &      17.8 &      0.908 &       1.11 &      0.985 &      0.997  \\
\rownumbervetoedfull & 5003670.3 & K04524.03 &      3.34 &      1.08 &       8.4 &      0.887 &       0.86 &      0.833 &      0.998  \\
\rownumbervetoedfull & 5621333.1 & K03341.01 &     27.10 &      2.30 &      14.7 &      0.865 &       1.04 &      0.994 &      0.997  \\
\rownumbervetoedfull & 7256914.2 & K04136.02 &      4.03 &      1.20 &      11.3 &      0.856 &       1.02 &      0.971 &      0.996  \\
\rownumbervetoedfull & 4851530.1 & K01884.01 &     23.08 &      2.07 &      17.6 &      0.846 &       1.21 &      0.969 &      0.996  \\
\rownumbervetoedfull & 4548011.2 & K04288.02 &      9.09 &      0.92 &      10.0 &      0.839 &       1.08 &      0.813 &      0.998  \\
\rownumbervetoedfull & 7375348.2 & K00266.02 &     47.74 &      1.79 &      14.1 &      0.795 &       0.93 &      0.991 &      0.998  \\
\rownumbervetoedfull & 12785320.2 & K00298.02 &     57.38 &      1.70 &      17.4 &      0.794 &       0.98 &      0.994 &      0.998  \\
\rownumbervetoedfull & 5211199.2 & K02158.02 &      6.68 &      2.12 &      10.1 &      0.777 &       0.98 &      0.996 &      0.998  \\
\rownumbervetoedfull & 11253711.1 & K01972.01 &     17.79 &      3.48 &      31.7 &      0.749 &       6.49 &      0.949 &      0.997  \\
\rownumbervetoedfull & 6527078.1 & K04657.01 &      7.58 &      0.64 &      10.2 &      0.740 &       0.92 &      0.985 &      0.997  \\
\rownumbervetoedfull & 5972334.3 & K00191.03 &      0.71 &      1.26 &      21.1 &      0.740 &       0.91 &      0.899 &      1.000  \\
\rownumbervetoedfull & 7673192.5 & K02722.05 &     16.53 &      1.24 &       7.8 &      0.710 &       1.05 &      0.984 &      1.000  \\
\rownumbervetoedfull & 5384713.2 & K03444.03 &      2.64 &      0.64 &      11.4 &      0.653 &       1.95 &      0.973 &      1.000  \\
\rownumbervetoedfull & 11135694.1 & K04896.02 &     49.54 &      1.84 &       9.3 &      0.643 &       1.04 &      0.983 &      0.996  \\
\rownumbervetoedfull & 6467363.1 & K02840.01 &      3.68 &      0.90 &      15.4 &      0.621 &       1.03 &      0.992 &      0.997  \\
\rownumbervetoedfull & 10471621.1 & K02554.02 &     10.27 &      1.07 &      13.8 &      0.618 &      -1.00 &      0.992 &      0.997  \\
\rownumbervetoedfull & 5384713.3 & K03444.01 &     12.67 &      0.84 &      10.1 &      0.614 &       1.95 &      0.994 &      1.000  \\
\rownumbervetoedfull & 10397751.5 & K02859.03 &      4.29 &      0.61 &       7.4 &      0.589 &       1.65 &      0.557 &      0.999  \\
\rownumbervetoedfull & 8838950.1 & K02421.01 &      2.27 &      0.62 &      13.1 &      0.585 &       3.57 &      0.892 &      0.991  \\
\rownumbervetoedfull & 11030475.2 & K02248.03 &      0.76 &      1.26 &      16.8 &      0.579 &      -1.00 &      0.921 &      0.999  \\
\rownumbervetoedfull & 6268648.4 & K01613.03 &     20.61 &      0.95 &       8.1 &      0.540 &      45.45 &      0.810 &      0.997  \\
\rownumbervetoedfull & 5531953.1 & K01681.01 &      6.94 &      1.18 &      18.4 &      0.540 &      -1.00 &      0.997 &      1.000  \\
\rownumbervetoedfull & 6527078.2 & K04657.02 &     10.43 &      0.74 &       8.9 &      0.520 &       0.92 &      0.969 &      0.996  \\
\rownumbervetoedfull & 4770174.1 & K02971.01 &      6.10 &      1.58 &      14.3 &      0.514 &       1.04 &      0.948 &      0.995  \\
\rownumbervetoedfull & 4851530.2 & K01884.02 &      4.78 &      1.89 &      14.3 &      0.461 &       1.21 &      0.987 &      0.997  \\
\rownumbervetoedfull & 11967788.2 & K04021.02 &      4.93 &      1.42 &      14.0 &      0.416 &      20.55 &      0.987 &      0.997  \\
\rownumbervetoedfull & 10875007.1 & K04149.01 &      9.55 &      1.72 &      13.9 &      0.350 &       2.05 &      0.988 &      0.997  \\
\rownumbervetoedfull & 8008067.3 & K00316.03 &      7.31 &      2.18 &      33.3 &      0.268 &       1.07 &      0.739 &      0.997  \\
\rownumbervetoedfull & 5542466.1 & K01590.01 &     12.89 &      2.36 &      18.7 &      0.099 &       5.07 &      0.980 &      1.000  \\
\bottomrule
\end{tabular}
\end{tabularx}
\end{threeparttable}
}
\end{table*}

\section{Caveats}
\label{sec:caveats}
Caveats related to the base classifier (\ExoMiner\ V1.2 in our case), or the multiplicity boost framework can affect the performance of the model and lead to erroneous validations. We have discussed the caveats related to \ExoMiner\ in~\cite{Valizadegan_2022_ExoMiner}. Here, we discuss those related to the multiplicity boost framework. 

The shape of the multiplicity mapping is affected by the form of the underlying classifier model used for multiplicity boost, i.e., logistic regression here, and largely dependent on the data set we use for training the model. The logistic regression makes certain assumptions that might not fully hold for the multiplicity data:
\begin{itemize}
\item Logistic regression requires that the independent variables are linearly related to the log odds. This assumption basically dictates the shape of the mapping between the original score and multiplicity score (Figure \ref{fig:multiplicity_curve}). Having a different assumption about the log odds will change the shape of the mapping even though the general behavior will stay the same. 
\end{itemize}

There are also caveats directly related to the multiplicity boost independent of the multiplicity classifier we use. This is mainly due to the fact that the multiplicity boost assumes that the candidates are not 1) FPs due to background objects and 2) FAs, as we discuss below:
\begin{itemize}
\item{FP due to background objects:} The multiplicity boost approach will fail for FPs due to background objects when there are multiple existing planets and unknown KOIs. One example of this situation is K02433.05. Even though K02433.05 is a background FP, the multiplicity approach boosts its very low score to 0.987 because there are three CPs and two unknown KOIs for this system. Fortunately, this does not pass the validation threshold. However, there could exist similar KOIs that pass the validation threshold. Our vetoing conditions mitigate this problem for most stars, though positional probability is not computed for all targets. For K02433.05, not only there are two FP flags, ``Stellar Eclipse Flag'' and ``Centroid Offset Flag' set, but also it has a low MES of 9.09.

Calculating the rate of misclassification for background FPs can provide great insights. However, this is not easy as we do not have the gold standard labels for background FPs. In order to provide an estimate, we use the KOI flag for centroid shift, i.e., ``Centroid Offset Flag,'' as the indicator for background FPs. There are a total of 1742 KOIs with this flag set. Out of these 1742 KOIs, \ExoMiner\ V1.2 correctly classifies 1731 KOIs as FPs (only 11 mistakes), resulting in a recall of 99.4\% for background FPs. Only one KOI out of these 11 misclassified cases, K00082.06, is in a multi-planet system. As mentioned in~\cite{Valizadegan_2022_ExoMiner} and reported in this manuscript, K00082.06 is incorrectly certified as FP in the Certified False Positive table. After applying the multiplicity boost model, a total of 14 KOIs, including the 11 KOIs originally misclassified by \ExoMiner\ V1.2, are misclassified. Except K00082.06, none of these 14 KOIs pass the validation threshold of 0.99 after the application of multiplicity boost. 

We also would like to mention that for a background FP to be incorrectly validated by our model, three conditions need to be satisfied: 1) \ExoMiner\ V1.2 gives a high enough score to that KOI; as we explained above, this is very rare, 2) that KOI should be around a multi-planet system with a high enough \ExoMiner\ score to get boosted above 0.99, and 3) all stability and vetoing conditions introduced in Section 5.3 and 5.4 fail. This can happen in practice but it is highly unlikely.
    
\item{FAs:} Given that the multiplicity boost works for planets and FPs, it might boost the score of an FA for scenarios that have multiple planets and unknown KOIs. We do not have any examples of this in the labeled KOIs but as we discussed in Section~\ref{sec:new_planets}, the unknown K00408.05 with multiplicity score $>0.99$ has the ``Not Transit-Like Flag'' on. It is also a KOI with MES $<10.5$. So our vetoing conditions such as MES $>10.5$ and KOI FP flags help for cases in this category, as well. 

In order to provide some insights regarding the chances of validating such observations, we would like to note that \ExoMiner\ V1.2 is highly accurate when it comes to the NTP instances. Out of 24474 NTPs in the \kepler\ data, there are only five with \ExoMiner score $>0.5$. This results in a 99.98\% recall for the NTP set. It is also highly accurate for FA objects in the KOI tables. In the CFP table~\cite{Bryson-2015-certifiedlist}, there are a total of 302 FAs out of which \ExoMiner\ V1.2 gives a score$>0.5$ to only two of them: K03226.01 with a score 0.84 and K02768.03 with a score 0.97, both of which are on systems with a single KOI (note that there is only one KOI for K02768 in Q1-Q17 DR25). Interestingly, K02768.03 is a transiting planet that has been confirmed in Q1-Q16 and incorrectly labeled as FA.

KOI 3226.01 was dispositioned as an FA by the Kepler False Positive Working Group based on Q1-Q16 data processing, and using that data this KOI violated the FA criteria by having large oscillations at the same period and with a similar shape as the transit signal~\cite{Bryson-2015-certifiedlist}.  In the final DR25 processing, however, these oscillations diminished so that they no longer violate the false alarm criterion.  Therefore we think of this KOI as not a clear case.

Overall, in order to have a validated FA KOI, it needs to satisfy the following three conditions: 1) \ExoMiner\ V1.2 gives a high score to that FA; this is very rare as we explained above, 2) the FA KOI should be around a system with enough other planets to be boosted to the validation threshold of 0.99 by the multiplicity boost framework, and 3) it needs to pass the vetoing and stability conditions explained in Section~\ref{sec:stability} and~\ref{sec:vetoes}. Having a FA that meets all these conditions is possible but unlikely.

\item{Application to TESS:} This same approach can also be beneficial when considering TESS exoplanet results. While the TESS pixel scale is $\sim 25$ times bigger in area than that for Kepler (21" vs 4" on a side), the average impact of background blends in TESS data and Kepler data are roughly comparable. This is due to the fact that TESS's exoplanet targets are $\sim~5$ magnitudes brighter than Kepler's, meaning that there are $\sim32$ times fewer background stars at a comparable delta-magnitude. Of course this average estimate breaks down in regions of high crowding (e.g., the Galactic plane or Galactic bulge), or in the cases of stars dimmer than Tmag$\sim$11, and results from these scenarios would require more careful checking and attention to other tests that provide information on background contamination, such as centroid and difference image analysis.
\end{itemize}

\section{Conclusions}
We introduced a new multiplicity boost framework that can be applied to any existing transit-signal classifier to boost its performance. Our framework does not require re-training or re-designing an existing classifier and can be applied to the output scores of a given classifier to improve its performance using the multiplicity information. We applied our framework to multiple state-of-the-art transit signal classifiers to demonstrate that multiplicity information improves their performance. Furthermore, we applied it to an improved version of \ExoMiner\ and validated 69 new exoplanets for systems with more than one candidate.

\section*{Acknowledgements}
This work is dedicated to the Women, Life, Freedom movement in Iran. We would like to symbolically name four of the newly validated exoplanets, K01358.01, K01358.02, K01358.03, and K01358.04, to Zan, Zendegi, Azadi\footnote{Persian words for woman, life, and freedom, respectively.}, and Iran, respectively.

Hamed Valizadegan and Miguel Martinho are supported through NASA NAMS contract NNA16BD14C, TESS GI Cycle 4 contract 80NSSC22K0184, and NASA ROSES XRP proposal 22-XRP22\_2-0173. Douglas Caldwell, and Joseph Twicken are supported through NASA Cooperative Agreement 80NSSC21M0079. We would like to thank multiple people who directly or indirectly contributed to this work. This paper includes data collected by the \kepler\ mission and obtained from the MAST data archive at the Space Telescope Science Institute (STScI). Funding for the \kepler\ mission was provided by the NASA Science Mission Directorate.  Resources supporting this work were provided by the NASA High-End Computing (HEC) Program through the NASA Advanced Supercomputing (NAS) Division at Ames Research Center for the production of the Kepler SOC data products and for training our deep learning model, \ExoMiner\ V1.2. This research has made use of the Exoplanet Follow-up Observation Program (ExoFOP; DOI: 10.26134/ExoFOP5) website, which is operated by the California Institute of Technology, under contract with NASA under the Exoplanet Exploration Program.


\bibliography{ExoPlanet_all}{}
\bibliographystyle{aasjournal}



\end{document}